\documentstyle[galley,epsf]{mn}
\title{A high-resolution spectral analysis of three carbon-enhanced metal-poor stars} 
\title[A high-resolution spectral analysis of three carbon-enhanced metal poor stars]
{A high-resolution spectral analysis of three carbon-enhanced metal-poor 
stars\thanks{Based on data collected at the Subaru Telescope, which is 
operated by the National Astronomical Observatory of Japan }}

\author[Aruna Goswami et al. ]{Aruna Goswami$^{1}$, Wako Aoki$^{2}$, 
    Timothy C. Beers$^{3}$, Norbert Christlieb$^{4,5}$, 
 \newauthor John E. Norris$^{6}$, Sean G. Ryan$^{7}$, Stelios Tsangarides$^{8}$  \\
    $^{1}$Indian Institute of Astrophysics, Koramangala, Bangalore 560034,
    India; aruna@iiap.res.in \\
    $^{2}$National Astronomical Observatory, Mitaka, Tokyo, 181-8588 Japan;
    aoki.wako@nao.ac.jp\\
    $^{3}$Department of Physics and Astronomy, CSCE: Center for the Study
    of Cosmic Evolution, and JINA: Joint Institute for Nuclear\\
    Astrophysics, Michigan State University, East Lansing, MI 48824-1116
    USA; beers@pa.msu.edu \\
    $^{4}$Hamburger Sternwarte, University of Hamburg, D-21029, Hamburg,
    Germany\\
    $^{5}$Department of Astronomy and Space Physics, Uppsala University,
    Box 515, SE-75120 Uppsala, Sweden; norbert@astro.uu.se\\
    $^{6}$Research School of Astronomy and Astrophysics, Australian
    National University, Canberra, ACT2611; jen@mso.anu.edu.au\\
    $^{7}$Centre for Astrophysics Research, STRI, University of
    Hertfordshire, College Lane, Hatfield, AL10 9AB, United Kingdom\\
    s.g.ryan@herts.ac.uk \\
    $^{8}$Department of Physics and Astronomy, Open University, Walton
    Hall, Milton Keynes, MK7 6AA, England, UK; \\
    s.tsangarides@open.ac.uk  \\
}    
\begin{document}

\date{ Accepted :  Received :  in original form :}

\pagerange{\pageref{firstpage}--\pageref{lastpage}} \pubyear{2006}

\maketitle

\label{firstpage}

\begin{abstract}

We present results of an analysis of high-resolution spectra ($R \sim
50\,000$), obtained with the Subaru Telescope High Dispersion Spectrograph,
of two Carbon-Enhanced Metal-Poor (CEMP) stars selected from the
Hamburg/ESO prism survey, HE~1305$+$0007 and HE~1152$-$0355, and of the
classical CH star HD~5223.  All of these stars have relatively low
effective temperatures (4000--4750\,K) and high carbon abundances, which
results in the presence of very strong molecular carbon bands in their
spectra. The stellar atmospheric parameters for these stars indicate that
they all have surface gravities consistent with a present location on the
red-giant branch, and metallicities of ${\rm [Fe/H]} = -2.0$
(HE~1305$+$0007, HD~5223) and ${\rm [Fe/H]} = -1.3$ (HE~1152$-$0355). In
addition to their large enhancements of carbon (${\rm [C/Fe]} = +1.8$,
$+1.6$ and $+0.6$. respectively), all three stars exhibit strong
enhancements of the s-process elements relative to iron.

HE~1305$+$0007 exhibits a large enhancement of the 3rd-peak s-process
element lead, with ${\rm [Pb/Fe]} = +2.37$, as well as a high abundance of
the r-process element europium, ${\rm [Eu/Fe]} = +1.97$. The 2nd-peak
s-process elements Ba, La, Ce, Nd, and Sm are found to be more enhanced
than the 1st-peak s-process elements Zr, Sr and Y. Thus, HE~1305$+$0007
joins the growing class of the so-called ``Lead Stars'', and also the class
of objects that exhibit the presence of both r-process and s-process
elements, the CEMP-r/s stars. The large enhancement of n-capture elements
exhibited by HE~1152$-$0355 and HD~5223 are more consistent with the
abundance patterns generally noticed in CH stars, essentially arising from
pure s-process nucleosynthesis. The elemental abundance distributions
observed in these stars are discussed in the light of existing theories of
CH star formation, as well as the suggested formation scenarios of the
CEMP-r/s group.

\end{abstract}


\section{Introduction}

Over the past few decades, detailed studies of the harvest of metal-poor
stars from the HK survey of Beers and colleagues (Beers, Preston, \&
Shectman 1985; Beers, Preston, \& Shectman 1992; Beers 1999) and the
Hamburg/ESO Survey (HES) of Christlieb and collaborators (Christlieb 2003)
have revealed them to possess a wide variety of elemental abundance
patterns. Among the at-first surprising results was the large fraction of
very metal-poor (${\rm [Fe/H]} < -2.0$) stars with large overabundances of carbon
relative to iron. Lucatello et al. (2006) estimates the fraction of
carbon-enhanced metal-poor (CEMP) stars, defined as 
having ${\rm [C/Fe]} > +1.0$,
to be on the order of 20\% to 25\% of very metal-poor stars, once the
likely dilution of carbon within their atmospheres is taken into
account. Subsequent high-resolution spectroscopic studies of a number of
CEMP stars (Norris et al. 1997a 1997b, 2002; Bonifacio et al. 1998; Hill et
al. 2000; Aoki et al. 2002; Aoki et al. 2006) has revealed that the CEMP
stars can be further subdivided into several classes. According to the
suggested taxonomy of Beers \& Christlieb (2005), these include: (a) the
CEMP-r stars, which exhibit the presence of strong enhancements of
r-process elements, (b) the CEMP-s stars, which exhibit the presence of
strongly enhanced s-process elements, (c) the CEMP-r/s stars, which
apparently have contributions from both of these neutron-capture sources,
and (d) the CEMP-no stars, which exhibit no enhancements of neutron-capture
elements.  Numerous questions remain as to the nucleosynthetic histories
and astrophysical sites associated with the production of these classes of
carbon-enhanced metal-poor stars. It already seems clear that a single
mechanism for the production of the enhanced carbon in these stars would be
unlikely to lead to such a diversity of heavy-element abundance patterns.

In order to obtain insight into the possible origins of the CEMP stars, and
to place additional constraints on the nucleosynthesis of s-process and
r-process elements at low metallicity, it is thus necessary to conduct
high-resolution observations of as many types of the CEMP stars as
possible. As carbon plays a central role in the nucleosynthesis reactions
that drive the post main-sequence evolution of stars, one might also obtain
clues into the evolutionary processes by studying the range of carbon
abundances and carbon isotope ratios exhibited by CEMP stars.

Christlieb et al. (2001) selected a large sample of carbon-enhanced stars
from the HES on the basis of the presence of molecular (e.g, CH, CN, and
C$_2$) bands on the original prism plates, without placing any additional
selection on their metallicities. Several teams have been obtaining low- to
medium-resolution spectroscopic follow-up of these stars in order to
investigate their nature in more detail. Marsteller et al. (in preparation)
will report on medium-resolution spectroscopy of over 350 stars from the
Christlieb et al. (2001) sample. Goswami (2005) reported spectral
classifications for 91 of the Christlieb et al. (2001) stars based on
low-resolution spectroscopy ($R = {\lambda}$/${\delta\lambda} \sim$
1330). In this study some 30\% of the sample were shown to be likely CEMP
stars. We have undertaken a high-resolution study of these CEMP candidates
(with additional targets from Goswami 2006, in preparation) in order to
obtain confirmation of their identification with this class, as well as to
carry out a comprehensive study of their heavy-element abundance
distributions.

This paper reports the first results of our high-resolution spectroscopic
campaign, and discusses two candidate CEMP stars from the Goswami (2005,
2006 in preparation) study, HE~1305$+$0007 and HE~1152$-$0355.  First
abundance results for a number of heavy elements in the classical CH star
HD~5223 are also presented.

In section 2 we present details of the observations that have been
conducted, and describe the low-resolution spectra of our program
stars. Section 3 presents $BVRIJHK$ photometry of these stars, and
discusses our estimates of their effective temperatures based on this
information. The procedures we have adopted for estimation of the stellar
atmospheric parameters are discussed in section 4.  The elemental abundance
results are presented in section 5. Results and discussions of our present
study, and consideration of existing theories for the nucleosynthesis
history of CEMP-s and CEMP-r/s stars are presented in section 6.  Section 7
provides some conclusions.

\section{Observation, Data Reduction, and Radial Velocities}

The low-resolution spectra of our program HES stars 
were obtained with the 2m Himalayan Chandra Telescope (HCT) at the Indian
Astronomical Observatory (IAO), Mt. Saraswati, Digpa-ratsa Ri, Hanle, India
during January 2005. The spectra of two classical CH stars, HD~5223 and
HD~209621, were also  obtained with the same telescope. 
These spectra cover a wavelength range from about $3800$\,{\AA}
to $6800$\, {\AA}, with a spectral resolving power $R
=\lambda/\delta\lambda \sim$ 1330. For each star two spectra were taken,
each of 15 minutes duration; the spectra were then combined to increase the
final signal-to-noise ratio. Observations of a Th-Ar hollow cathode lamp
taken immediately before and after the stellar exposures provided the
wavelength calibration. Standard spectroscopic reductions (e.g., flat
fields, bias subtraction, extraction, and wavelength calibration) were
carried out using the IRAF\footnote{IRAF is distributed by the National
Optical Astronomical Observatories, which is operated by the Association
for Universities for Research in Astronomy, Inc., under contract to the
National Science Foundation} spectroscopic reduction package.

High-resolution spectroscopic observations of our program stars were
carried out with the High Dispersion Spectrograph (HDS) of the 8.2m Subaru
Telescope (Noguchi et al. 2002) on 25 May, 2003. Each object spectrum was
taken with a 10 minute exposure having a resolving power of $R \sim$
50\,000. The observed bandpass ran from about $4020$\,{\AA} to
$6775$\,{\AA}, with a gap of about $75$\,{\AA}, from $5335$\,{\AA} to
$5410$\,{\AA}, due to the physical spacing of the CCD detectors. These data
were also reduced, in the standard fashion, using IRAF.

\begin{figure*}
\epsfxsize=14truecm
 \epsffile{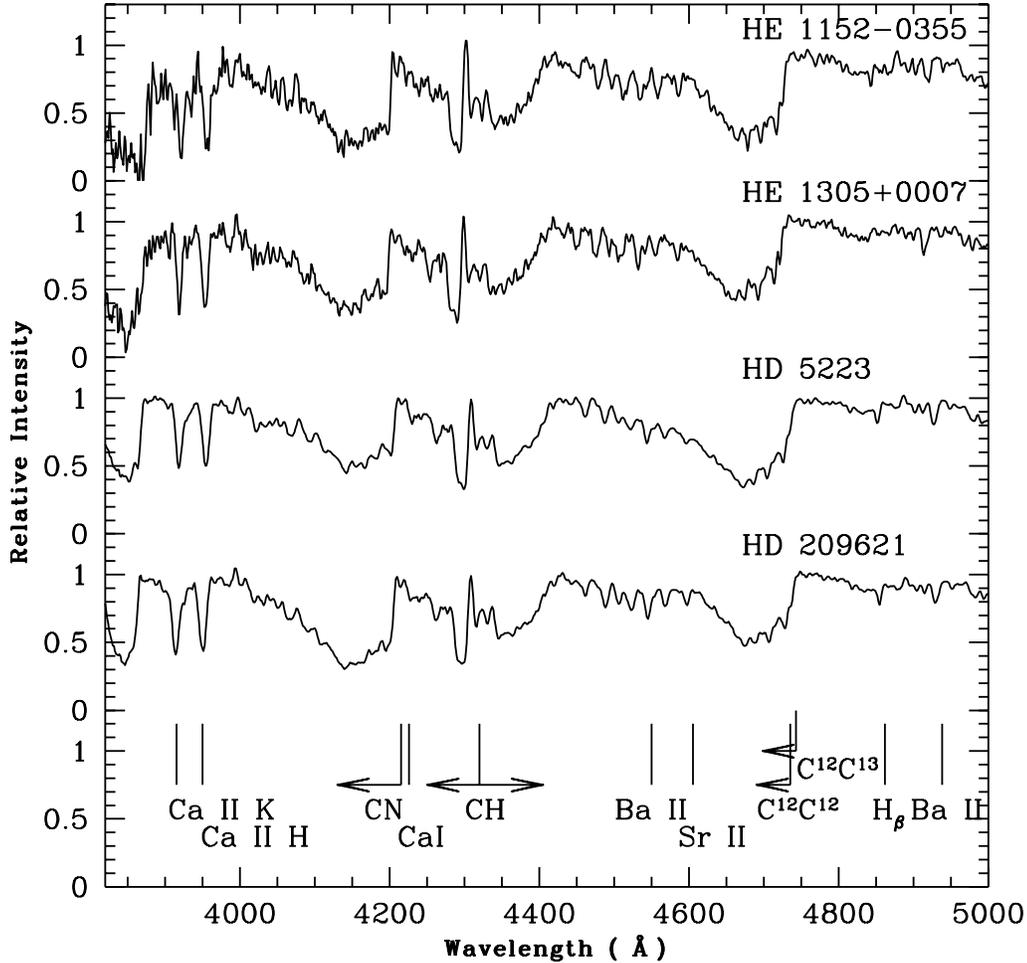}
\caption{A comparison of the low-resolution spectra of HE~1305$+$0007 and
HE~1152$-$0355 in the wavelength region $3850$\,{\AA} to $5000$\,{\AA}. The
spectra of the well-known CH stars HD~5223 and HD~209621 are also plotted
for comparison. Prominent atomic and molecular features are indicated.}
\label{Figure 1}
\end{figure*}

\subsection{Description of the spectra}

Figure 1 shows the low-resolution spectra of the two HES stars in the
wavelength region $3850$\,{\AA} to $5000$\,{\AA}. The spectra of two
well-known CH stars HD~5223 and HD~209621, obtained at the same resolving
power, are also plotted for comparison. Following Goswami (2005), we note
that the two HES stars are potential CEMP (CH) candidate stars. The G band
of CH is noticeably strong and appears to be almost of equal strength in
all the spectra. The secondary P-branch head around $4342$\,{\AA} is
distinctly seen in the HES stellar spectra, as well as in HD~5223 and
HD~209621. The Ca~I feature at $4226$\,{\AA} is weakly noticed in the two
HES stellar spectra. Narrow atomic lines are blended with contributions
from molecular bands. As in the case of the two HD stars, H${\beta}$ and
Ba~II at $4554$\,{\AA} are two clearly noticable features in the HES
stellar spectra.

The high-resolution spectra of HE~1305$+$0007, HE~1152$-$0355, and HD~5223
are typical of the complex combination of molecular and atomic absorption
features in their cool atmospheres. The optical spectra of the stars are
characterized by very closely-spaced molecular absorption lines of CH, CN
and the Swan system of C$_{2}$. The continuum is obscured over essentially
the entire wavelength region, and the atomic absorption lines are
overwhelmed by the molecular transitions. Measurements of equivalent widths
are therefore not free from uncertainties, even in the case of relatively
clean lines. Nonetheless, we have identified a number of atomic features on
the spectra and measured their equivalent widths, which are used in the
present work to derive the elemental abundances using a model-atmosphere
analysis.

We have also searched for Li in the high resolution spectra of our three
program stars. A wide spread of Li abundances is known to exist among CEMP
stars (e.g., Norris, Ryan \& Beers 1997a). Although in giants Li is not
expected to currently exist on their stellar surfaces due to its likely
destruction from internal-mixing processes, it is still worthwhile to
search for its presence. If detected, a measure of Li abundance could help
understand the formation mechanisms of these objects. Hill et al (2000)
have derived a Li abundance for CS~22948-027, (a star very similar to
HE~1305$+$0007) of log$ {\epsilon}$(Li) = +0.3, a value much lower than the
primordial Li abundance. This is an indication of some internal mixing
affecting the surface abundances. Unfortunately, this feature could not be
detected in the spectra of our stars.

\subsection{Radial velocities}

The radial velocities of our program stars were measured using several
unblended lines.  Estimated heliocentric radial velicities $v_{\rm r}$ of
the program stars are shown in Table 1.
No previous estimates of radial velocities for HE~1305$+$0007 and
HE~1152$-$0355 are available in the literature. It is thus presently not
known whether these two stars are radial velocity variables. HD~5223, on
the other hand, is known to be a radial velocity variable; McClure \&
Woodsworth (1990) studied its radial velocity variations in detail and
reported that $v_{\rm r}$ of HD 5223 ranges from $-$230 to $-$250 km
s$^{-1}$. The SIMBAD database lists a mean radial velocity of $v_{\rm r} =
-232$ km s$^{-1}$ for this star.

\section {Photometry}

Optical broadband $BVRI$ photometry has been obtained for our program stars
(Beers et al. 2006). These data are reported, along with near-IR $JHK$
photometry from 2MASS (Skrutskie et al. 2006), in Table 2. Estimates of the
line-of-sight reddening values, $E(B-V)$, for the HES stars listed in Table
2 are obtained from the Schlegel, Finkbeiner, \& Davis (1998) dust
maps. The reddening estimate for CS~22948-027 is $E(B-V) = 0.00$, as
obtained from the Burstein and Heiles (1982) maps, and $E(B-V) = 0.04$ in
the case of HD~5223, as obtained from Sleivyte \& Bartkevicius (1990).

{\footnotesize
\begin{table*}

{\bf Table 1:  Heliocentric Radial velicities  v$_{r}$  of the program stars}\\

\begin{tabular}{ l c c c c  }
  &   &   &  &    \\

\hline

Stars   & $v_{\rm r}$ km s$^{-1}$  & HJD    &  $v_{\rm r}$ km s$^{-1}$ \\
        & our estimation     &        & from literature    \\ 
\hline
 HE~1305$+$0007& +217.8 ${\pm}$ 1.5   & 2452784.95121 & ---        \\
 HE~1152$-$0355& +431.3 ${\pm}$ 1.5 & 2452784.83629 & ---        \\
HD~5223      &$-$244.9 ${\pm}$ 1.5 & 2453544.12049 & $-$232     \\
\hline
\end{tabular}
\end{table*}
}

{\footnotesize
\begin{table*}

{\bf Table 2: Photometric parameters for program and comparison stars}\\

\begin{tabular}{ l c c c c c c c c c c  c }
  &   &   &  &  &   &  &     &   &  \\

\hline

Stars   &RA(2000)  & Dec(2000)     &  V   & $B-V$   &$V-R$   & $V-I$   &$E(B-V)$
&  $J$    &$ H$  & $K_{s}$\\
\hline
 HE~1305$+$0007& 13 08 03.84  & $-$00 08 47.3  &12.22& 1.459 &0.682 & 1.152 & 0.038 & 10.247 & 9.753 & 9.600\\ 
 HE~1152$-$0355& 11 55 06.10 & $-$04 12 24.0  &  11.43& 2.459 &0.816 & 1.194 & 0.026 &  9.339 & 8.665 & 8.429\\
HD~5223      &00 54 13.61  &+24 04 01.5   &8.47 & 1.43  &  ---    & ---      & 0.04  &  6.372 & 5.845 & 5.673 \\
CS~22948-027 &21 37 45.80 & $-$39 27 22.0  & 12.65 & 1.12  &0.5   & 0.90  & 0.00  & 10.979 & 10.534 & 10.427\\
     
\hline

\end{tabular}
\end{table*}
}

\subsection{Effective temperatures from photometry}

Estimates of the effective temperatures of the two HES stars in our program
have been determined using the temperature calibrations derived by Alonso
et al.  (1996), which relate $T_{\rm eff}$ with various optical and near-IR
colours.  Alonso et al. estimate an external uncertainty in the temperature
calculation using this method of $\sim 90$\,K.

The Alonso calibrations of $B-V$, $V-R$, $V-I$, $R-I$ and $V-K$ require
colours in the Johnson system, and the calibrations of the IR colours
$J-H$, and $J-K$ in the TCS system (the photometric system at the 1.54~m
Carlos Sanchez telescope in Tenerife; Arribas \& Martinez-Roger 1987). To
obtain the $V-K$ colour in the Johnson system we first transformed the
K$_{s}$ 2MASS magnitude to the TCS system. Then, using eqs (6) and (7) of
Alonso et al. (1994), K is transformed to the Johnson system from the TCS
system. In order to transform the 2MASS colours $J-H$ and $J-K_S$ onto the
TCS system we first transformed the 2MASS colours to CIT colours (Cutri et
al. 2003), and then from CIT to the TCS system (Alonso et al. 1994).

Estimation of the $T_{\rm eff}$ from the $T_{\rm eff}$ - $(J-H)$ and
$T_{\rm eff}$ - $(V-K)$ relations also involves a metallicity ([Fe/H])
term. We have estimated the $T_{\rm eff}$ of the stars at several
metallicities. The estimated temperatures, along with the adopted
metallicities, are listed in Table 3.

The broadband $B-V$ colour is often used for the determination of $T_{\rm
eff}$, however, the $B-V$ colour of a star with strong molecular carbon
absorption features depends not only on $T_{\rm eff}$, but also on the
metallicity of the star and on the strength of its molecular carbon
absorption features, due to the effect of CH molecular absorption in the B
band. For this reason, we have not used the empirical $T_{\rm eff}$ scale
for the $B-V$ colour indices. The temperatures listed in Table 3 span a
wide range; while $J-K$, $J-H$, and $V-K$ colours provide similar
temperatures, the $V-I$ colour predict higher temperatures.

{\footnotesize
\begin{table*}
{\bf Table 3: Estimated effective temperatures ($T_{\rm eff}$) from 
semi-empirical relations }\\
\begin{tabular}{lcccccc}
\hline
Star Names  &$T_{\rm eff}$ &$T_{\rm eff}$ &$T_{\rm eff}$ & $T_{\rm eff}$  & $T_{\rm eff}$ &\\
            &  $(J-K)$        &  $(J-H)$    &  $(V-K)$    & $(V-R)$ &     
$(V-I)$   &    \\
\hline
HE~1305$+$0007&      4636 & 4697.2 ($-$1.5) & 4497.8 ($-$1.5)  & 4949.9 ($-$1.5) & 5043.4    &      \\
            &           & 4714.3 ($-$2.0) & 4376.2 ($-$2.0)  & 4931.2 ($-$2.0) &---         &    \\ 
HE~1152$-$0355&    3756 & 3985.9 ($-$1.0) & 4133.4 ($-$1.0)  & 4529.8 ($-$1.0) &  4963.2    &    \\
            &           & 4002.3 ($-$1.5) & 4083.8 ($-$1.5)  & 4451.8 ($-$1.5) &---         &    \\ 
HD 5223     &      4360 & 4539.0 ($-$1.0) & 4298.5 ($-$1.0)  &  ---            &---          &  \\
            &           & 4555.9 ($-$1.5) & 4253.9 ($-$1.5)  &  ---            &---          &  \\
            &           & 4572.9 ($-$2.0) & 4213.8 ($-$2.0)  &  ---            &---          &   \\
CS 22948$-$027&    4892 & 4903.8 ($-$1.0) & 4840.5 ($-$1.0)  & 5784.8 ($-$1.0) &  5629.8   &     \\
            &           & 4920.7 ($-$1.5) & 4815.8 ($-$1.5)  & 5758.4 ($-$1.5) &---        &     \\
            &           & 4937.7 ($-$2.0) & 4796.1 ($-$2.0)  & 5755.1 ($-$2.0) &---          &    \\
\hline

\end{tabular}
 \\
The numbers inside the parentheses indicate the adopted metallicities [Fe/H] \\
\end{table*}
}

\section{Stellar atmospheric parameters }

A selection of Fe~I and Fe~II lines (Table A), covering a range in
excitation potential (0.0-4.0 eV) and equivalent widths (20-160 m\AA), was
used in a routine procedure to determine the stellar atmospheric parameters
-- the effective temperature $T_{\rm eff}$, the surface gravity log $g$,
the microturbulence $V_{\rm t}$, and the metallicity [Fe/H]. For
HE~1152$-$0355, a few stronger lines were also used, as the number of lines
falling in the above range of equivalent widths is very small. Nineteen
Fe~I lines in HE~1305$+$0007, eighteen Fe I~lines in HE~1152$-$0355, and
twenty-four Fe~I lines (and four Fe~II lines) in HD~5223 were found to be
useful for our analysis. In Table B, we present the measured equivalent
widths for other elements. In this Table, the star HE 1152-0355 is not
included, due to the severe contamination by molecular absorption features
in this object, which prevented reasonable fits of gaussian line profiles
to be obtained. This star is also the coolest among the three. Throughout
our analysis we have assumed that local thermodynamic equilibrium
conditions apply. Model atmospheres were selected from the Kurucz grid of
model atmospheres with no convective overshooting, and from those that are
newly computed models with better opacities and abundances. These models
are available at {\tt http://cfaku5.cfa.harvard.edu/}, labelled with the
suffix ``odfnew''. For our analysis, the excitation potentials and
oscillator strengths of the lines were taken from various sources,
including the Vienna Atomic Line Database ({\tt
http://ams.astro.univie.ac.at/vald/}) and also the Kurucz atomic line list
({\tt
http://cfa-www.harvard.edu/\-amdata/\-ampdata/\-kurucz23/\-sekur.html}),
Fuhr, Martin, \& Wiese (1988), Martin, Fuhr, \& Wiese (1988), and Lambert
et al. (1996). The $gf$ values for some elements were also taken from a
compilation assembled by R.E.  Luck (private communication); these are
based, when possible, on experimental determinations of high quality. We
have employed the latest (2002) version of MOOG, an LTE stellar
line-analysis program (Sneden 1973).

The microturbulent velocity is generally estimated at a given effective
temperature by demanding that there should be no dependence of the derived
Fe~I abundance upon the equivalent widths of Fe I~lines. However, in the
present case, the number of usable lines is small. The wavelength regions
around the lines at 4863, 4892, 5753 and $5806$\,{\AA} that are used in our
analysis are significantly affected by C$_{2}$ and CN molecular
features. It is therefore possible that the equivalent widths of these
lines are overestimated due to contamination from molecular features. This
in turn would affect the determination of microturbulent velocity. Taking
this into consideration, and following the discussions of Vanture (1992)
and McWilliam et al. (1995) ( $V_{\rm t}$ in cool giants with log\,$g \le$
2.0 are in general ${\ge}$ 2 km s$^{-1}$), we have adopted a value of
$V_{\rm t}$ = 2 km s$^{-1}$ for both of the HES stars and for HD~5223.

Estimates of effective temperatures were then obtained by the method of
excitation balance, forcing the slope of the abundances from Fe~I lines
versus excitation potential to be near zero. The temperature estimates in
Table 3 provided a preliminary temperature check for the program stars;
model atmospheres corresponding to these temperatures were used initially,
and the adopted effective temperatures were then obtained by an iterative
process using the method of excitation balance. In Figure 2, we show the
iron abundances of two examples, HE~1305$+$0007 and HD~5223, for individual
Fe~I and Fe~II lines as a function of each line's equivalent width and the
lower excitation potential. If we rely on the Fe~I excitation equilibrium,
with an adopted microturbulance of 2 km s$^{-1}$, we estimate the effective
temperatures for HE~1305$+$0007 and HE~1152$-$0355 to be 4750 K and 4000 K,
respectively. For HD~5223, the estimated $T_{\rm eff}$ is 4500~K. Using the
Fe~I/Fe~II ionisation equilibrium, the surface gravities of HE~1305$+$0007,
HE~1152$-$0355 and HD~5223 are obtained to be log $g$ = 2.0, 1.0 and 1.0,
respectively. The derived metallicities are ${\rm [Fe/H]} = -2.0$, $-$1.3 and
$-$2.0, respectively (Table 4).

Vanture (1992) reported a higher metallicity, ${\rm [Fe/H]} = -0.9$, for HD~5223,
even though his estimates of $T_{\rm eff}$ $\sim$ 4550 K and log\,$g$
$\sim$ 1.3 are very similar to those we have adopted. To verify our
estimations, we have determined the atmospheric parameters for the star
CS~22948-027 (a star very similar to HE~1305$+$0007, see section 6) and
compared with those given by other authors. We obtained similar results to
those reported by Barbuy et al. (2005). They estimated $T_{\rm eff} =
4800$K, log\,$g$ = 1.8, $V_{\rm t} = 1.5$ and ${\rm [Fe/H]} = -2.46$, very close
to our present estimates (Table 4).

\section{Abundance Analysis}

There are numerous features, from a variety of elements, identified in our
program star spectra (Table B). However, throughout the spectral range the
line blending is so severe that for most elements a standard abundance
analysis procedure based on the equivalent widths could not be applied, and
the abundances must be derived from spectrum-synthesis calculations. A fit
of the synthetic spectrum in the wavelength region 5170 to $5190$\,{\AA},
obtained using the atmospheric parameters listed in Table 4, is illustrated
in Figure 3. This region is relatively free from contamination by molecular
features. The method of spectrum synthesis is also applied to lines that
are affected by hyperfine splitting. For the spectrum synthesis of lines,
we find that a projected velocity $v$sin$i$ of 5.0 ${\pm}$ 2 km s$^{-1}$
matches the widths of the observed features with those of the synthetic
spectra.  This value  represents the sum of macroscopic broadening, and
does not necessarily indicate significant rotation of the star.  In Table 5
we present abundances derived from spectrum-synthesis calculations. In this
Table we have listed not only the abundance log $\epsilon$(X), and [X/H],
but also [X/Fe]. In computing the quantity [X/Fe], we have used the
Fe~I-based abundance for elemental abundances derived from neutral lines
and the Fe~II-based abundance for elemental abundances derived from ionized
lines.  Local thermodynamic equilibrium was assumed for all of the
spectrum-synthesis calculations. We have used the latest version of MOOG
for spectrum synthesis.  The line lists for each region that is synthesized
are taken from the Kurucz atomic line list ({\tt
http://cfa-www.harvard.edu/\-amdata/\-ampdata/\-kurucz23/\-sekur.html)} and
from the
Vienna Atomic Line Database ({\tt http://ams.astro.univie.ac.at/vald/}).
Reference solar abundances for the various elemental species were adopted from
Asplund, Grevesse \& Sauval (2005). The log\,$gf$ values for atomic lines were
adopted from Fuhr et al. (1988) and Martin et al. (1988), whenever available,
and also from a compilation of $gf$ values by R. E. Luck (private communication).
The heavy neutron-capture elements lines and corresponding log\,$gf$ values
given by Sneden et al. (1996) and Lawler et al. (2001) were also consulted.

For HE~1305$+$0007, we have determined the abundances of C, Na, Mg, Ca, Sc,
Ti, Mn, Ni, and among the heavy elements, Sr, Y, Zr, Ba, La, Ce, Pr, Nd,
Sm, Eu, and Pb. Abundances of several heavy s-process elements, in addition
to the carbon abundance, are also determined for HE~1152$-$0355. For
HD~5223 we could determine the abundances of C, Na, Mg, Ca, Ti, Ni, Sr, Y,
Zr, Ba, La, Ce, Nd, Sm, and Pb. Details of the behaviour of the elemental
abundance patterns for our program stars are discussed below.

\subsection{Carbon}

The carbon (C) abundance is derived from spectral synthesis of the C$_{2}$
Swan 0-1 band around $5635$\,{\AA}. A synthetic spectrum, derived using a
carbon abundance of log ${\epsilon}$(C) = 8.2 ${\pm}$ 0.3, combined with
the appropriate model atmosphere, shows a good match to the depth of the
observed spectrum of HE~1305$+$0007. Relative to the solar photospheric C
abundance, C is strongly enhanced in HE~1305$+$0007 (${\rm [C/Fe]} = +1.84$ ). A
similar procedure returns a C abundance of log ${\epsilon}$(C) = 7.7 $\pm$
0.1 for HE~1152$-$0355 (${\rm [C/Fe]} = +0.58$), and log ${\epsilon}$(C) = 7.9
$\pm$ 0.2 for HD~5223 (${\rm [C/Fe]} = +1.57$).

Carbon abundances derived from the spectrum synthesis of the CH G-band are
not considered. This feature is severely saturated in our objects, is quite
insensitive to the carbon abundance, and hence likely to return uncertain
values.  Spectrum synthesis of the C$_{2}$ Swan 0-1 band around
$5635$\,{\AA}\ is, however, found to provide the most reasonable
estimates. Spectrum-synthesis fits for HE~1305$+$0007, HE~1152$-$0355, and
HD~5223 are shown in Figure 4.

\subsection{The odd-Z elements Na and Al }

The sodium (Na) abundance in HE~1305$+$0007 is calculated from the
resonance doublet -- the Na I D lines at $5890$\,{\AA} and
$5896$\,{\AA}. These resonance lines are sensitive to non-LTE effects
(Baum\"uller \& Gehren 1997; Baum\"uller et al. 1998; Cayrel et
al. 2004). The derived abundance from an LTE analysis displays an
overabundance with respect to Fe of ${\rm [Na/Fe]} = +0.26$, based on the Na~I
D$_{2}$ line, and an overabundance ${\rm [Na/Fe]} = +0.43$, based on the Na~I
D$_{1}$ line. This abundance may be therefore slightly overestimated (the
non-LTE case is not considered here). As a result of the severe line
asymmetry of the Na~I resonance doublet, due to molecular contamination, we
could not estimate the Na abundance in the star HE~1152$-$0355. In
CS~22948-027, an upper limit of the Na abundance was estimated to be
${\rm [Na/Fe]} < +0.57$ by Barbuy et al. (2005). This value is slightly higher
than what we have obtained for HE~1305$+$0007, a star very similar to this
object. For HD~5223, the Na abundance is derived using the Na~I D lines,
and found to be ${\rm [Na/Fe]} = +0.46$.  While Na exhibits a mild overabundance in
HE~1305$+$0007 and CS~22948-027, it is interesting to note that the excess
of Na is quite significant in CS~29497-034, a star which is otherwise very
similar to both HE~1305$+$0007 and CS~22948-027.  This star also has the
lowest metallicity amongst these three.

The aluminum (Al) lines are severely blended in the spectra of our program
stars, and could not be used for abundance determination. The Al line at
$3961.5$\,{\AA}, which is generally used for Al abundance determination in
extremely metal-poor stars, lies outside the wavelength range of our
spectra.

\subsection{The ${\alpha}$-elements  Mg, Si, Ca, and Ti }

Several lines due to ${\alpha}$-elements were identified in the spectra of
the two HES stars. However, in order to avoid contamination problems from
the strong molecular carbon features, we have selected the cleanest line of
each element, and derived the abundance from spectrum-synthesis
calculations. Hence, in many cases, the abundance is derived from a
single-line spectrum-synthesis calculation.

The Mg abundance in HE~1305$+$0007 is derived from the synthesis of the
Mg~I line at 5172.68 {\AA}. The predicted line profile with our adopted Mg
abundance (log $\epsilon$(Mg) = 5.75) fits the observed line profile in
HE~1305$+$0007 quite well. Magnesium is found to exhibit the usual
halo-star overabundance, [Mg/Fe] $\sim +0.25$. In HE~1305$+$0007 and
HD~5223, the Mg abundances are practically as expected for a halo star with
${\rm [Fe/H]} = -2.0$; i.e., showing the classical enhancement of the
${\alpha}$-elements (see Goswami \& Prantzos 2000, Figure 7).  Spectrum
synthesis of this line in HE~1152-0355, however, returns an almost solar
value of ${\rm [Mg/Fe]} = -0.01$. The calcium (Ca) abundance is derived from
spectrum synthesis of the Ca~I line at $6102.7$\,{\AA}. Calcium also
exhibits a normal halo overabundance of ${\rm [Ca/Fe]} = +0.13$ and +0.1 in
HE~1305$+$0007 and HD~5223, respectively. The titanium (Ti) abundance
derived from the synthesis of the Ti~II line at $4805.085$\,{\AA} shows a
marked overabundance of ${\rm [Ti/Fe]} {\sim} +0.8$ in HE~1305$+$0007. For HD
5223, Ti is overabundant with respect to iron by +0.5 dex.

Due to the presence of severe line blending, the ${\alpha}$-element
abundances, with the exception of Ti, could not be derived in the star
HE~1152$-$0355. As in the case of HE~1305$+$0007, the Ti abundance in
HE~1152$-$0355 is also derived from a spectrum synthesis of the Ti~II line
at $4805.085$\,{\AA}; we obtain ${\rm [Ti/Fe]} \sim +0.5$ for HE~1152-0355.

\subsection{The iron-peak elements Mn and Ni}

For HE~1305$+$0007, a spectrum synthesis of the Mn~I line at $5432.56$\,
{\AA} returned an abundance estimate of log $\epsilon$(Mn) = 3.5; [Mn/Fe]
is found to be almost solar in this star, with ${\rm [Mn/Fe]} = +0.14$. A Mn
abundance could not be determined for HE~1152$-$0355 and HD~5223.

\begin{figure*}
\epsfxsize=10truecm
 \epsffile{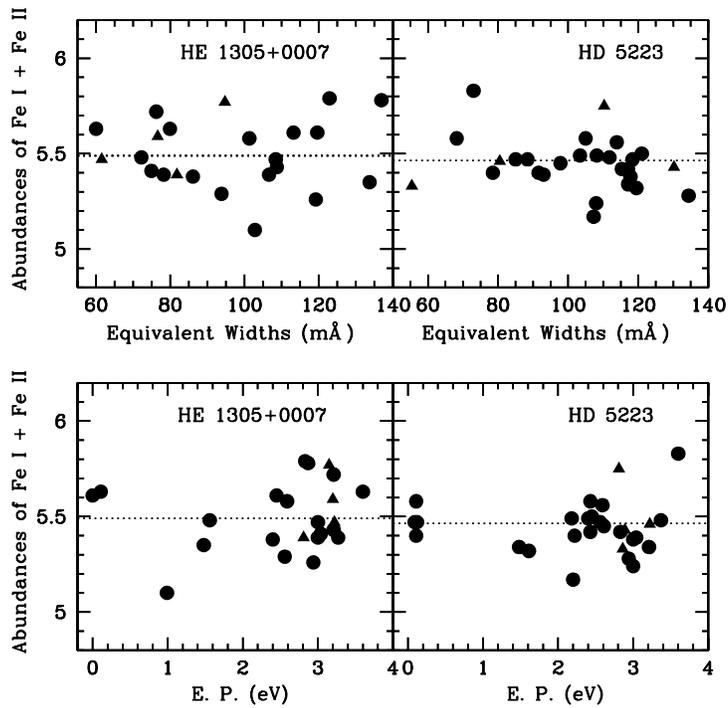}
\caption{ The iron abundance of HE~1305$+$0007 (left panels) and HD~5223
(right panels) are shown for individual Fe~I and Fe~II lines as a function
of the line's equivalent width (upper panels) and the lower excitation
potential (lower panels). The solid circles indicate Fe~I lines and the
solid triangles indicate Fe~II lines.  }
\label{Figure 2}
\end{figure*}

\begin{figure*}
\epsfxsize=10truecm
 \epsffile{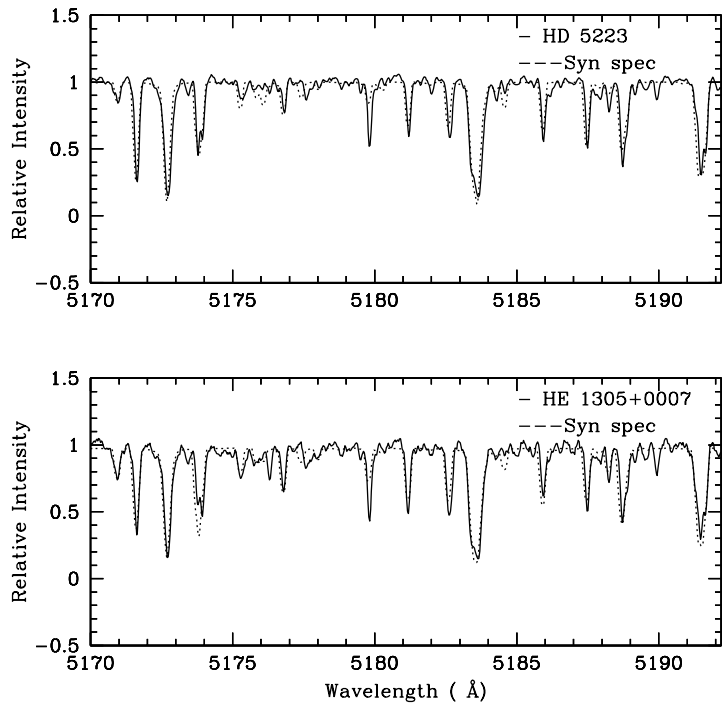}
\caption{ A fit of the synthetic spectrum (dotted curve) compared with the
observed spectrum (solid curve) of HD~5223 (upper panel) and HE~1305$+$0007
(lower panel) in the wavelength region 5170 to $5190$\,{\AA}. The synthetic
spectra are obtained using a model atmosphere corresponding to the adopted
parameters listed in Table 4. }
\label{Figure 3}
\end{figure*}
\clearpage

\begin{figure*}
\epsfxsize= 12truecm
 \epsffile{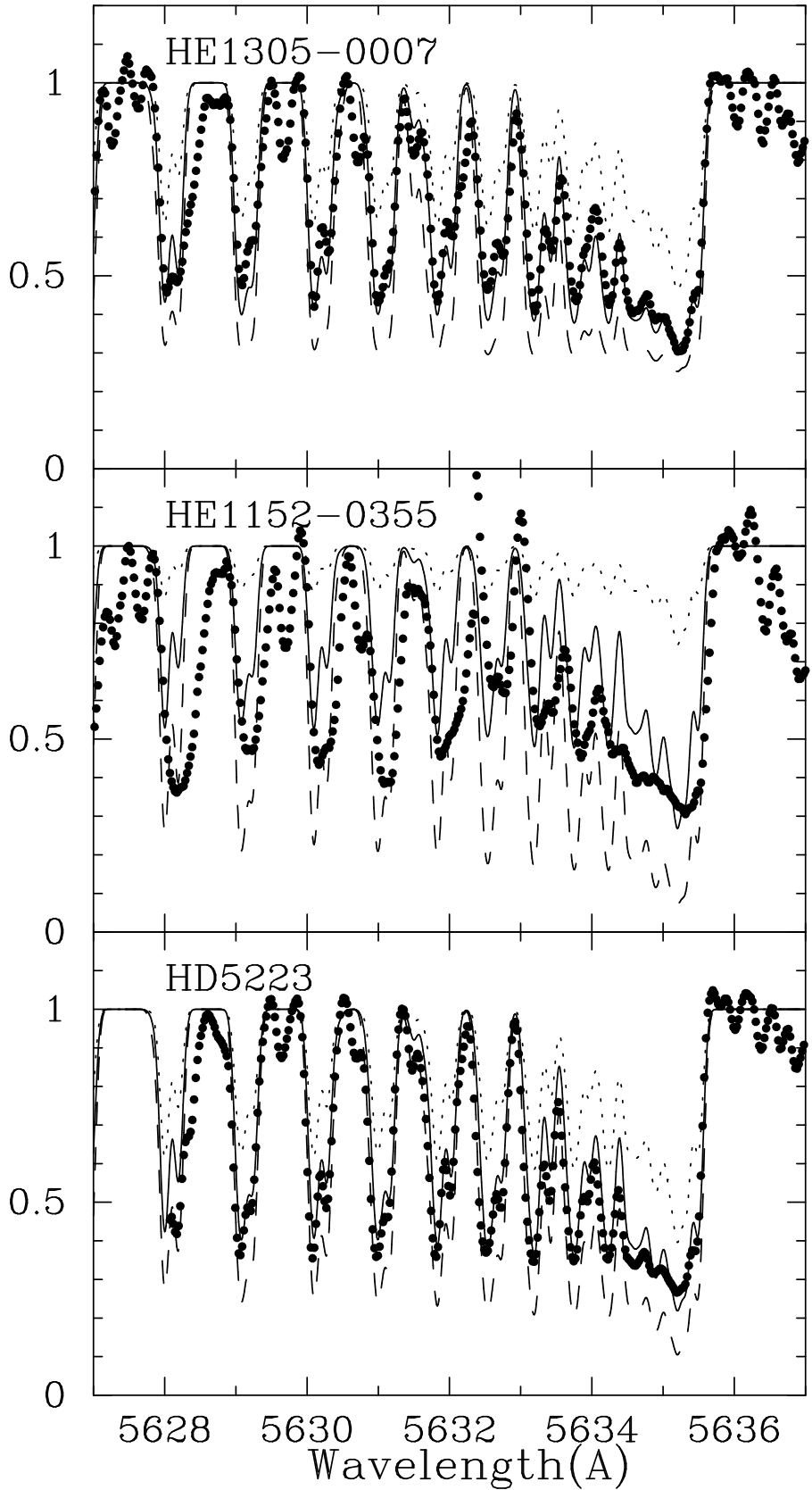}
\caption{A comparison of the fits of the synthetic spectra for log
${\epsilon}$(C) = 8.2 $\pm$ 0.3 (top), 7.7 $\pm$ 0.1 (middle) and 7.9 $\pm$
0.2 (bottom) (represented by solid line, dash and dotted curve,
respectively) with the observed spectrum of HE~1305$+$0007, HE~1152$-$0355,
and HD 5223, respectively (shown by solid circles) in the wavelength region
around $5635$\,{\AA} of the C$_{2}$ Swan 0-1 bands. Carbon abundances of
log ${\epsilon}$(C) = 8.2, 7.7, and 7.9 provide the best fit (solid line)
of the synthetic spectra with the observed spectra of HE~1305$+$0007,
HE~1152$-$0355, and HD~5223, respectively. }
\label{Figure 4 }
\end{figure*}
\clearpage

\begin{figure*}
\epsfxsize=  9truecm
 \epsffile{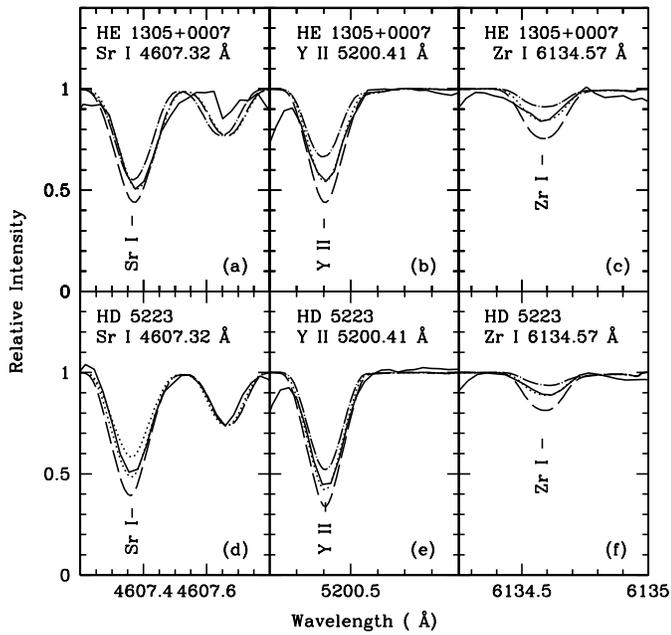}
\caption{Spectral-synthesis fits of absorption lines arising from the light
s-process elements Sr, Y, and Zr, obtained with the elemental abundances
listed in Table 5.  The dotted lines indicate the synthesized spectra and
the solid lines indicate the observed line profiles. Two alternative
synthetic spectra for ${\Delta}$[X/Fe] = +0.3 (long dash) and
${\Delta}$[X/Fe] = $-$0.3 (dot-dash) are  shown to demonstrate the
sensitivity of the line strength to the abundances. }
\label{Figure 5 }
\end{figure*}

\begin{figure*}
\epsfxsize= 9truecm
 \epsffile{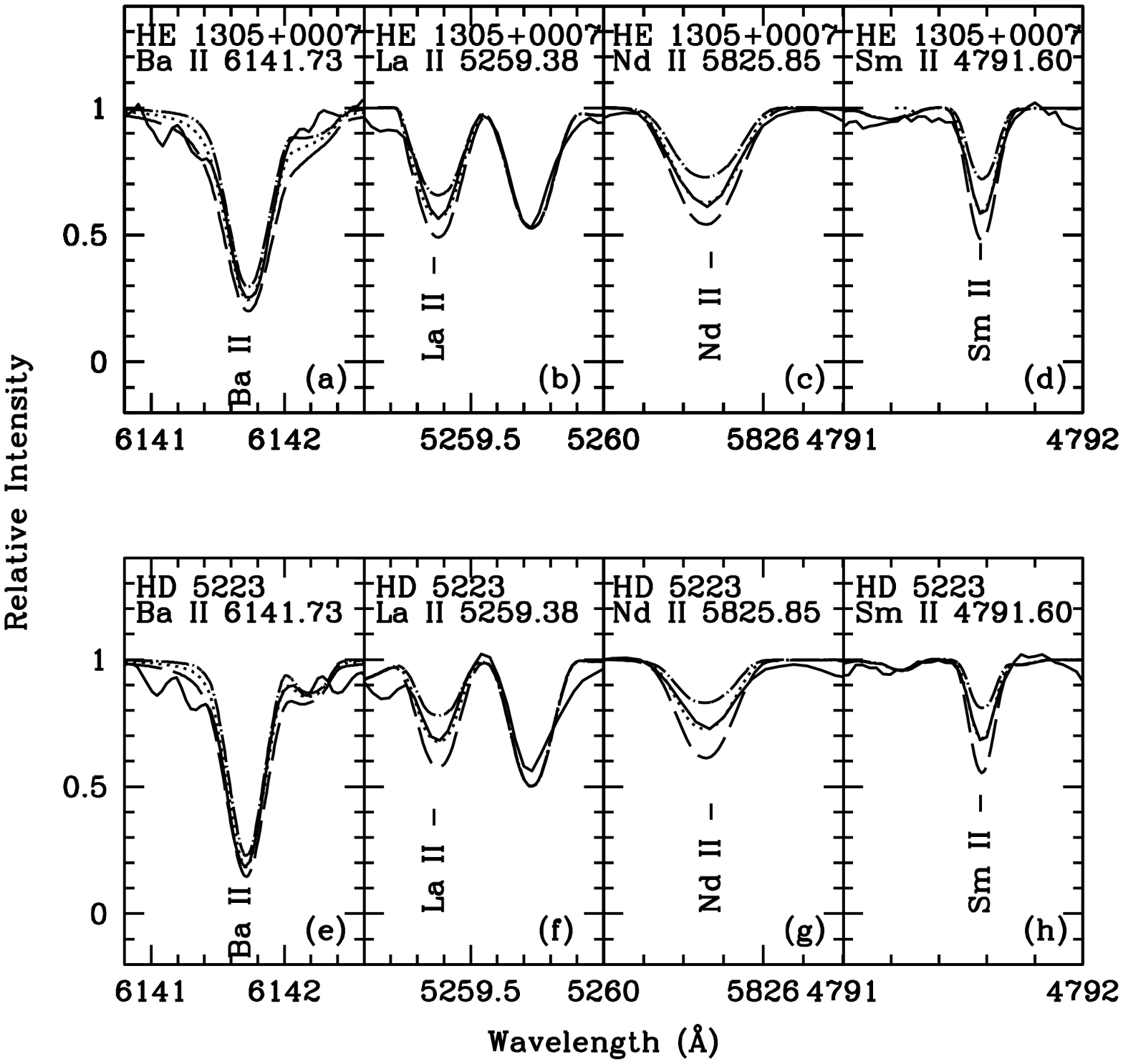}
\caption{Spectral-synthesis fits of absorption lines arising from the
heavy s-process elements Ba, La, Nd, and Sm, obtained with the respective
elemental abundances listed in Table 5. The dotted lines indicate the
synthesised spectra and the solid lines indicate the observed line
profiles.  Two alternative synthetic spectra for ${\Delta}$[X/Fe] = +0.3
(long dash) and ${\Delta}$[X/Fe] = $-$0.3 (dot-dash) are shown to
demonstrate the sensitivity of the line strength to the abundances.  }
\label{Figure 6 }
\end{figure*}
 \clearpage

\begin{figure*}
\epsfxsize= 9truecm
 \epsffile{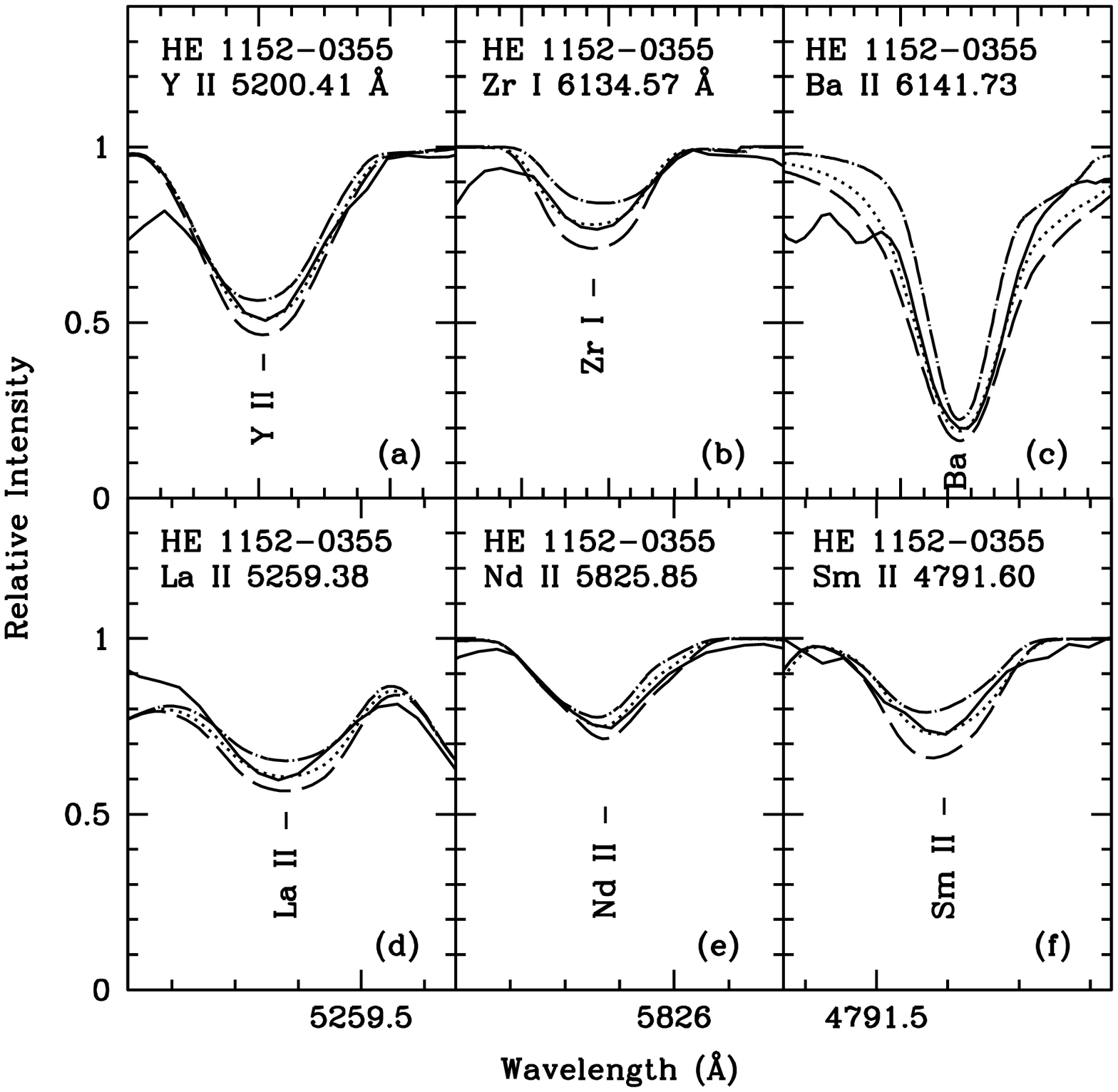}
\caption{Spectral-synthesis fits of absorption lines arising from the heavy
s-process elements Y, Zr, Ba, La, Nd, and Sm, obtained with the respective
elemental abundances listed in Table 5. The dotted lines indicate the
synthesised spectra and the solid lines indicate the observed line
profiles.  Two alternative synthetic spectra for ${\Delta}$[X/Fe] = +0.2
(long dash) and ${\Delta}$[X/Fe] = $-$0.2 (dot-dash) are  shown to
demonstrate the sensitivity of the line strength to the abundances.  }
\label{Figure 7 }
\end{figure*}

\begin{figure*}
\epsfxsize=10truecm
 \epsffile{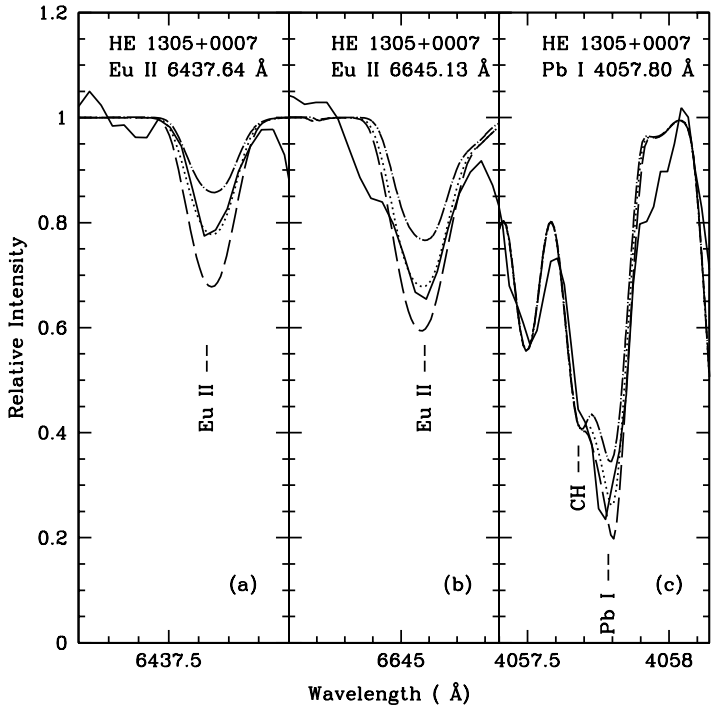}
\caption{Spectral-synthesis fits of the two Eu~II lines at $6437.64$\,{\AA}
and $6645.13$\,{\AA}, obtained with an europium abundance of log
$\epsilon$(Eu) = 0.50. The spectral-synthesis fit of the Pb~I line at
$4057.80$\,{\AA} is obtained with log $\epsilon$(Pb) = 2.38. The dotted
lines indicate the synthesized spectra and the solid lines indicate the
observed line profiles.  Two alternative synthetic spectra for
${\Delta}$[X/Fe] = +0.3 (long dash) and ${\Delta}$[X/Fe] = $-$0.3
(dot-dash) are  shown to demonstrate the sensitivity of the line
strength to abundances.  }
\label{Figure 8 }
\end{figure*}

{\footnotesize
\begin{table*}
{\bf Table 4: Derived  atmospheric parameters }\\
\begin{tabular}{lcccccc}
\hline
Star Names  &$T_{\rm eff}$ &log\,$g$ & $V_{\rm t}$ km s$^{-1}$ &[Fe I/H]& [Fe II/H] &\\
            &           &         &                    &         &     &    \\
\hline

HE~1305$+$0007   &  4750   & 2.0   & 2.0 & $-$2.03   & $-$1.99 &  \\
HE~1152$-$0355 &  4000   & 1.0   & 2.0 & $-$1.27   & $-$1.30 &  \\
HD 5223        &  4500   & 1.0   & 2.0 & $-$2.06   & $-$2.04 &   \\
CS 22948-027   &  4750   & 1.5   & 2.0 & $-$2.50   & $-$2.40 &  \\
\hline

\end{tabular}
\end{table*}
}

{\footnotesize
\begin{table*}

{Table 5: Chemical compositions }

\begin{tabular}{ l c c c c c c c  c c c c c }
\hline

Element& $Z$&Solar$^{a}$&HE~1305$+$0007&  & &HE~1152$-$0355& & &HD~5223& & \\ 
       &    & log ${\epsilon}$ &log ${\epsilon}$ &[X/H]&[X/Fe]&log ${\epsilon}$&[X/H]&[X/Fe]&log ${\epsilon}$ &[X/H]&[X/Fe]\\
\hline
C I         &  6& 8.39& 8.20 & $-$0.19 &+1.84 & 7.7 & $-$0.69&+0.58   & 7.9 & $-$0.49&+1.57 &\\
Na I D$_{2}$& 11& 6.17& 4.40 & $-$1.77 &+0.26 & --- &  ---   &  ---   & 4.57& $-$1.60&+0.46 &\\
Na I D$_{1}$& 11& 6.17& 4.57 & $-$1.60 &+0.43 & --- &  ---   &  ---   & 4.57& $-$1.60&+0.46 &\\
Mg I        & 12& 7.53& 5.75 & $-$1.78 &+0.25 & 6.25 &$-$1.28&$-$0.01 & 6.05& $-$1.48&+0.58 &  \\
Ca I        & 20& 6.31& 4.41 & $-$1.90 &+0.13 & --- &  ---   &  ---   & 4.35& $-$1.96&+0.10 &  \\
Sc II       & 21& 3.05& 1.15 & $-$1.90 &+0.09 & --- &  ---   &  ---   & --- &---    & ---     &   \\
Ti II       & 22& 4.9 & 3.70 & $-$1.20 &+0.79 & 4.15& $-$0.75& +0.52  & 3.35 & $-$1.55&+0.49 & \\
Mn I        & 25& 5.39& 3.50 & $-$1.89 &+0.14 & --- &  ---   &  ---   & --- & ---  &  ---    &    \\ 
Fe I        & 26& 7.45& 5.42$\pm 0.18$ & $-$2.03 & ---  & 6.18$\pm 0.27$& $-$1.27&    ---    & 5.39$\pm 0.13$& $-$2.06& ---  &   \\
Fe II       & 26& 7.45& 5.46$\pm 0.17$ & $-$1.99 & ---  & 6.15$\pm 0.24$& $-$1.30&    ---    & 5.41$\pm 0.18$& $-$2.04& ---  &  \\
Ni I        & 28& 6.23& 3.95 & $-$2.28 &-0.25 & --- &  ---   &  ---   & 3.70& $-$2.53&$-$0.47 &  \\
Sr I        & 38& 2.92& 1.75 & $-$1.17 &+0.86 & --- &  ---   &  ---   & 2.25& $-$0.67&+1.39 &  \\
Y II        & 39& 2.21& 0.95 & $-$1.26 &+0.73 & 1.05& $-$1.16& +0.14  & 0.80& $-$1.41&+0.63 & \\
Zr  I       & 40& 2.59& 2.65 & +0.06   &+2.09 & 1.32& $-$1.27&  0.0   & 2.10& $-$0.49&+1.57 & \\
Ba II       & 56& 2.17& 2.50 & +0.33   &+2.32 & 2.45& +0.28  & +1.58  & 1.95& $-$0.22&+1.82 & \\
La II       & 57& 1.13& 1.70 & +0.57   &+2.56 & 1.40& +0.27  & +1.57  & 0.85& $-$0.28&+1.76 &\\
Ce II       & 58& 1.58& 2.12 & +0.54   &+2.53 & --- &  ---   &  ---   & 1.75& $-$0.17&+1.87 &\\
Pr II       & 59& 0.71& 1.10 & +0.39   &+2.38 & --- &  ---   &  ---   & --- & ---    & ---     &  \\
Nd II       & 60& 1.45& 2.05 & +0.60   &+2.59 & 0.58& $-$0.87& +0.43  & 0.95& $-$0.50&+1.54 &     \\
Sm II       & 62& 1.01& 1.62 & +0.61   &+2.60 & 0.58& $-$0.43& +0.87  & 0.65& $-$0.36&+1.68 & \\
Eu II       & 63& 0.52& 0.50 & $-$0.02 &+1.97 & --- &  ---   &  ---   & --- & ---    & ---     & \\
Pb I        & 82& 2.00& 2.38 & +0.38   &+2.37 & --- &  ---   &  ---   & 2.15& +0.15  &+2.21$^{*}$ & \\
\hline
\end{tabular}
\\
$^{a}$ Asplund, Grevesse \& Sauval   (2005); $^{*}$ indicates an upper limit  \\
\end{table*}
}

Nickel abundances in our study are derived from a spectrum synthesis of the
Ni~I line at $6314.66$\,{\AA}. Nickel exhibits a mild underabundance of
${\rm [Ni/Fe]} = -0.25$ in HE~1305$+$0007, and an underabundance 
of ${\rm [Ni/Fe]} = -0.47$ in HD~5223.

For very metal-poor stars, a general decrease of the ratios of Mn and Cr,
relative to Fe, was initially found by McWilliam et al. (1995) and Ryan,
Norris, \& Beers (1996). The observational scatter in these trends was
lowered considerably by the more recent higher-quality data of Cayrel et
al. (2004).  Our measured abundances for both of these species fit well
with the Cayrel et al. trends.

\subsection{The light s-process elements Sr, Y, and Zr}

Abundances of the light s-process elements strontium (Sr), yttrium (Y), and
zirconium (Zr) are also determined from spectrum-synthesis
calculations. The Sr abundance is derived from spectrum synthesis of the
Sr~I line at $4607.327$~{\AA}, the Y abundance from the Y~II line at
$5200.41$\,{\AA}, and the Zr abundance is derived from the Zr~I line at
$6134.57$\,{\AA}. Examples are shown in Figure 5. In HE~1305$+$0007, Sr, Y,
and Zr exhibit overabundances of ${\rm [Sr/Fe]} = +0.86$, ${\rm [Y/Fe]} = +0.73$, 
and ${\rm [Zr/Fe]} = +2.09$, respectively. In HE~1152$-$0355, Y and Zr are found to be
almost solar. A spectrum synthesis calculation of Sr line could not be done
being highly blended . In HD~5223, Sr, Y, and Zr exhibit overabundances of
${\rm [Sr/Fe]} = +1.39$, ${\rm [Y/Fe]} = +0.63$, and ${\rm [Zr/Fe]} = +1.57$, 
respectively.

\subsection{The heavy $n$-capture elements:  Ba, La, Ce, Pr, Nd, Sm, Eu, Pb}

As in the case of the other elements, the abundances of barium (Ba),
lanthanum (La), cerium (Ce), neodymium (Nd), and samarium (Sm) are also
determined using spectrum-synthesis calculations (examples are shown in
Figure 6 and Figure 7).  The derived abundances for these elements are
found to be overabundant with respect to Fe in all three of our program
stars. In the case of HE~1305$+$0007, abundances of a few more heavy
elements, such as Pr, Eu, and Pb, are also determined (Figure 8).

We have not used any hyperfine-splitting corrections for Ce, Nd, and Y
lines, since they are very weak, and the abundances are not expected to be
affected by hyperfine splitting (McWilliam et al. 1995).

In the case of Ba the hyperfine-splitting (HFS) corrections depend on the
$r/s$ fraction assumed to have contributed to the enrichment of the
star. This element has both odd and even isotopes. The odd isotopes are
mainly produced by the r-process and have a broad HFS, while the even
isotopes are mainly produced by the s-process, and exhibit no HFS. However,
it was shown by Sneden et al. (1996) that this issue is important for the
Ba lines at $4554$\,{\AA} and $4934$\,{\AA}, and unimportant for the three
other Ba lines at 5854, 6142, and $6497$\,{\AA}, which are generally used
for abundance determination. We have used the red Ba~II line at
$6141.73$\,{\AA} for the abundance determination in our stars. The HFS
splitting of this line is $\sim$1/5 of the Ba~4554\,{\AA} splitting and
$\sim$1/3 of the thermal line width, and hence we have neglected HFS
corrections. Ba is found to exhibit a marked overabundance of ${\rm [Ba/Fe]} =
+2.32$ in HE~1305$+$0007, ${\rm [Ba/Fe]} = +1.58$ in HE~1152$-$0355, and 
${\rm [Ba/Fe]} = +1.82$ in HD~5223.

Lanthanum (La) abundances are derived from a spectrum-synthesis calculation
of the La~II line at $5259.38$\,{\AA}, with atomic data taken from Lawler
et al.  (2001). Similar to Ba, La also exhibits an overabundance of
 ${\rm [La/Fe]} = +2.56$ in HE~1305$+$0007, ${\rm [La/Fe]} = +1.57$ in 
HE~1152$-$0355, and ${\rm [La/Fe]} = +1.76$ in HD~5223.  Neodymium abundances 
are derived using the Nd~II line at
$5825.85$\,{\AA}.  Neodymium is also found to be highly overabundant in
HE~1305$+$0007, with ${\rm [Nd/Fe]} = +2.59$, whereas, in HE~1152$-$0355 this
element is only mildly overabundant, with ${\rm [Nd/Fe]} = +0.43$. In HD~5223,
${\rm [Nd/Fe]} = +1.54$.

Samarium (Sm) abundances are derived from the Sm~II line at
$4790.60$\,{\AA}.  In HE~1305$+$0007 , Sm exhibits an overabundance of
${\rm [Sm/Fe]} = +2.60$, while in HE~1152$-$0355 and HD~5223, 
overabundances of ${\rm [Sm/Fe]} = +0.87$  and ${\rm [Sm/Fe]} = +1.68$ 
are found, respectively.

In the spectrum of HE~1305$+$0007 we could also determine the europium (Eu)
abundance. Eu is usually considered to be produced almost solely by the
r-process in solar-system material. The main lines of this element that are
generally used in abundance analysis are the Eu~II lines at 4129.7,
4205.05, 6437.64 and $6645.13$\,{\AA}. The blue Eu~II lines at 4129.7 and
$4205.05$\, {\AA} are severely blended with strong molecular features in
this star, and could not be used for abundance analysis. The abundance of
Eu in HE~1305$+$0007 is therefore determined from the red lines at
$6437.64$\,{\AA} and $6645.13$\,{\AA}, using spectrum-synthesis
calculations. Both of the lines return an abundance of log $\epsilon$(Eu) =
0.50, indicating that Eu is highly overabundant in HE~1305$+$0007 
(${\rm [Eu/Fe]} = +1.97$). Illustrations of the spectral synthesis for 
the Eu lines at 6437.64 and $6645.13$\,{\AA} are shown in Figure 8.

Abundances of cerium (Ce) and praeseodymium (Pr) are derived from
spectrum-synthesis calculations of the Ce~II line at $5274.23$\,{\AA} and the
Pr~II line at $5259.7$\,{\AA}, respectively. Both of these elements exhibit
large overabundances in HE~1305$+$0007; ${\rm [Ce/Fe]} = +2.53$ and
 ${\rm [Pr/Fe]} = +2.38$, respectively. In HD~5223, ${\rm [Ce/Fe]} = +1.87$, 
hence it is also strongly overabundant.

We have also used spectrum-synthesis calculations to determine the
abundances of Pb in HE~1305$+$0007, using the Pb I line at
$4057.8$\,{\AA}. This line is strongly affected by molecular absorption of
CH (as indicated in Figure 8). CH lines are included in our spectrum
synthesis calculation of Pb abundance. The best spectrum-synthesis fit with
the observed line profile is obtained with a lead abundance of log
$\epsilon$(Pb) = 2.38, as shown in Figure 8, an overabundance with respect
to iron of ${\rm [Pb/Fe]} = +2.37$. This line could not be detected in 
the spectrum of HE~1152$-$0355. Spectrum synthesis of this line returns 
a lead abundance of log~${\epsilon}$(Pb) = 2.15 in HD~5223, with a 
corresponding overabundance of ${\rm [Pb/Fe]} = +2.21$.

\subsection{Error Analysis}

Errors in the derived abundances arise mainly from two sources, random
errors as well as systematic errors arising from uncertainties in our
adopted atmospheric parameters. Errors arising due to uncertainties in line
parameters, such as the adopted $gf$ values and equivalent width
measurements, are usually random and cause line-to-line scatter in derived
abundances for a given species.  The random errors are minimized by
employing as many usable lines as possible for a given element. In deriving
the Fe abundances we made use of 19, 18, and 24 Fe I lines and 4, 2, and 4
Fe II lines in HE~1305$+$0007, HE~1152-0355 and HD~5223, respectively. The
derived standard deviation ${\sigma}$ is defined by ${\sigma}^{2}$ =
[${\Sigma}(\chi_{i} - {\chi})^{2}/(N-1)$], where $N$ is the number of lines
used. The values of ${\sigma}$ computed from the Fe I lines are ${\pm}$0.18
dex, ${\pm}$0.27 dex and ${\pm}$0.13 dex in HE~1305$+$0007, HE~1152-0355,
and HD~5223, respectively. The corresponding values for these three stars
calculated for Fe II lines are ${\pm}$0.17 dex, ${\pm}$0.24 dex, and
${\pm}$0.18 dex, respectively. Errors arising from the random uncertainties
are computed when the number of clean lines ($N$) measured is ${\ge}$ 2 for
a given element. The computed errors for Fe I and Fe II are listed in Table
5.

Uncertainties in the adopted model atmospheric parameters may also
introduce errors in derived abundances. The accuracy of the atmospheric
parameters was estimated by computing a set of Fe I lines for pairs of
models with (a) the same gravity and microturbulence velocity but different
temperatures (b) with the same temperature and gravity but different
microturbulent velocities and (c) with the same temperature and
microturbulent velocity but different gravities. A comparison of the
variations in the computed equivalent widths for the three cases with the
accuracy of equivalent width measurements allowed us to estimate the
uncertainties in the determination of the atmospheric parameters. The
(conservative) uncertainties in the estimated $T_{\rm eff}$ is $\pm 250K$,
in surface gravity log\, g $\pm 0.25$ dex, and in microturbulent velocity
$V_{\rm t}$ $\pm$ 0.25 km s$^{-1}$.  Errors in the abundances arising from
the errors in atmospheric parameters are not a simple sum of the errors due
to the individual parameters. These parameters interact with one another
and a change in one may cause a shift in another. The net effect on the
derived mean abundances should be considerably less, because we employ the
principle of consistency wherein the lines with a large range of excitation
potentials, equivalent widths, and different ionization states should lead
us to the same value of abundances.

In Table 6, we take HE~1305$+$0007 as an example, and derive differential
abundances of elements with respect to those obtained from the adopted
model ($T_{\rm eff}$ = 4750 K, log\,g = 2.0, and $V_{\rm t}$ = 2.0 km
s$^{-1}$). These errors are estimated by varying $T_{\rm eff}$ by $\pm$ 250
K, log\,g by 0.25 dex, and $V_{\rm t}$ by 0.25 km s$^{-1}$ in the adopted
stellar atmosphere models of HE~1305$+$0007. Computation of abundance
errors for the other two stars yields similar results.

The abundances of all other elements listed in Table~5 are derived by
spectrum synthesis calculation. In these cases we have visually estimated
the fitting errors. Our estimated fitting errors range between 0.1 dex and
0.3 dex. We adopt these fitting errors as estimates of the random errors
associated with the derived elemental abundances. In Figures 4 through 8,
we have shown synthetic spectra for the adopted element abundances compared
with the synthetic spectra due to two other possible abundances with
${\pm}$0.2 dex or ${\pm}$0.3 dex (see figure captions) differences with
respect to our adopted abundances.

{\footnotesize
\begin{table*}

{Table 6: Estimation of errors. Differential abundances  ${\Delta}$log${\epsilon}$ 
of elements with respect to those derived using the adopted model with
 $T_{\rm eff}$ = 4750 K, log\,g = 2.0, and $V_{\rm t}$ = 2.0 km s$^{-1}$ for HE~1305$+$0007}

\begin{tabular}{ l c c c c c c c c c c c c }
\hline
       & & ${\Delta}$$T_{\rm eff}$&  &${\Delta}T_{\rm eff}$&  &${\Delta}$log\,${g}$&  &
${\Delta}$log\,${g}$& &${\Delta}$$V_{\rm t}$ &  &${\Delta}$$V_{\rm t}$ \\

Element &      &  +250   &          & $-$250   &         & +0.25    &          &  $-$0.25   &       & +0.25    &        & $-$0.25   \\
\hline
 Na   &      & +0.35 &   & $-$0.37 &   &$-$0.05 &   &  $-$0.04 &    & $-$0.14 &    &+0.03   \\
 Mg   &      & +0.30 &   & $-$0.50 &   &$-$0.08 &   &   +0.12 &    & $-$0.15 &    &+0.05   \\
 Ca   &      & +0.34 &   & $-$0.24 &   & +0.03  &   &   +0.06 &    &  +0.05 &     &+0.11   \\
 Sc   &      & +0.17 &   & $-$0.15 &   & +0.10  &   &  $-$0.10 &    & $-$0.05 &    &+0.04     \\
 Ti   &      &$-$0.02&   & $-$0.05 &   & +0.08  &   &  $-$0.10 &    & $-$0.10 &    &+0.08   \\
 Mn   &      & +0.40 &   & $-$0.55 &   &$-$0.05 &   &  $-$0.03 &    & $-$0.10 &    &$-$0.06   \\
 Fe I &      & +0.36 &   & $-$0.36 &   &$-$0.02 &   &  $-$0.08 &    & $-$0.10 &    &+0.13   \\
 Fe II&      &$-$0.06&   &  +0.13 &    & +0.10 &    &  $-$0.17 &    & $-$0.07 &    &+0.12   \\
 Ni   &      & +0.34 &   & $-$0.28 &   & +0.03  &   &   +0.02 &    &   +0.01 &    &+0.05   \\
 Sr   &      & +0.30 &   & $-$0.35 &   & +0.10  &   &    +0.04 &    & $-$0.04 &    &+0.04   \\
 Y    &      & +0.07 &   & $-$0.05 &   & +0.10  &   &  $-$0.05 &    & $-$0.05 &    &+0.07   \\
 Zr   &      & +0.40 &   & $-$0.50 &   & $-$0.01&   &   +0.02 &    & $-$0.01 &    &+0.02   \\
 Ba   &      & +0.18 &   & $-$0.20 &   & +0.04  &   &  $-$0.05 &    & $-$0.10 &    &+0.10   \\
 La   &      & +0.10 &   & $-$0.10 &   & +0.08  &   &  $-$0.10 &    & $-$0.10 &    &+0.10   \\
 Ce   &      & +0.10 &   & $-$0.05 &   & +0.09  &   &  $-$0.09 &    & $-$0.10 &    &+0.12   \\
 Pr   &      & +0.10 &   & $-$0.10 &   & +0.08  &   &  $-$0.09 &    & $-$0.14 &    &+0.12    \\
 Nd   &      & +0.05 &   & $-$0.10 &    & +0.03  &    & $-$0.10 &    & $-$0.10 &    & +0.06    \\
 Sm   &      & +0.13 &   & $-$0.10 &    & +0.10  &    & $-$0.09 &    & $-$0.05 &    & +0.06    \\
 Eu   &      &$-$0.05 &   & $-$0.12 &    & +0.06  &    & $-$0.15 &    & $-$0.12 &    &$-$0.11   \\
 Pb   &      & +0.12 &   & $-$0.28 &    & +0.03  &    & $-$0.12 &    & $-$0.25 &    &$-$0.15   \\
\hline
\end{tabular}
\end{table*}
}

\section{ Results and Interpretations}

We have reported a high-resolution spectroscopic analysis of the CEMP stars
HE~1305$+$0007 and HE~1152$-$0355, which belong to the somewhat lesser
studied group of objects of cool CEMP stars. A detailed analysis of
high-resolution spectroscopy for HD~5223, a well-known classical CH star
(used by Goswami 2005, 2006 as a comparison star), has also been reported.

Our analysis confirms that HD~5223 and the two HES stars are all highly
carbon-enhanced metal-poor stars.  One characteristic feature of CH
(CEMP-s) stars is that the heavy 2nd -peak s-process elements, such as Ba,
La, Ce, Nd, and Sm, are more enhanced than the lighter 1st-peak s-process
elements  Sr, Y and Zr. All three of our program stars exhibit this
behaviour (Table 5).

Radial-velocity measurements (Table 1) indicate that all three objects are
high-velocity stars, and likely to be members of the Galactic halo
population. While we have measured a radial velocity of +217.8 km s$^{-1}$
in the case of HE~1305$+$0007, for HE~1152-0355 we have measured +431.3 km
s$^{-1}$, almost twice as high.  Such a high velocity for halo stars is
rare but not exceptional; an inspection of high proper motion stars (Carney
et al.  1994) reveals three stars, namely G~64$-$12, G~241$-$4, and
G~233$-$27 with absolute velocities greater than 400 km s$^{-1}$. From the
high radial velocity of HE~1152-0355, one could speculate that the star
could be a runaway from a binary companion that exploded (in the context of
Qian and Wasserburg 2003 scenarios of s-processing occuring in an AGB
member of a binary system). However, such an idea requires a more detailed
study of a greater number of stars in order to reach a definitive
conclusion.  As mentioned before, HD~5223 is known to be a radial-velocity
variable, and is likely a binary system. For the two HES stars our
measurements are the first available accurate radial-velocity measurements,
so we are unable to comment on their likely binarity.

{\footnotesize
\begin{table*}

{\bf Table A: List of Fe lines  }\\

\begin{tabular}{ c c c c c c c c }
  &      &   &            &           &              &             &     \\

\hline
         &    &           &          & HE~1305$+$0007 & HE~1152$-$0355 & HD 5223\\
W$_{lab}$& ID & EP$_{low}$& log${gf}$& Eq widths (m\AA\,)& Eq widths (m\AA\,)& Eq widths (m\AA\,)\\
  
\hline
 6593.880&Fe I&   2.4327  &  $-$2.420  &  ---   &  ---     &     68.2 \\ 
 6551.680&Fe I&   0.9901  &  $-$5.790  &  ---   &  74.7    & --- \\
 6518.380&Fe I&   2.8316  &  $-$2.750  &  ---   &  100.8   & --- \\ 
 6335.340&Fe I&   2.1979  &  $-$2.230  &  ---   &  159.9   & --- \\ 
 6252.550&Fe I&   2.4041  &  $-$1.687  &   86.1 &  ---     &    108.2  \\
 6230.726&Fe I&   2.5590  &  $-$1.281  &   93.8 &  208.0   &    118.4  \\ 
 6219.280&Fe I&   2.1979  &  $-$2.430  &  ---   &  206.8   & ---  \\ 
 6213.430&Fe I&   2.2227  &  $-$2.660  &  ---   &  186.2   & ---  \\
 6191.558&Fe I&   2.4330  &  $-$1.600  &  ---   &  195.5   &   117.1  \\
 6137.694&Fe I&   2.5881  &  $-$1.403  &  101.3 &  ---     &  113.9    \\ 
 6136.615&Fe I&   2.4530  &  $-$1.400  &  113.2 &  199.2   &  121   \\ 
 6065.480&Fe I&   2.6085  &  $-$1.530  &  ---   &  204.8   &   97.75    \\
 5853.180&Fe I&   1.4848  &  $-$5.280  &  ---   &  67.0    & ---   \\
 5586.760&Fe I&   3.3683  &  $-$0.210  &  ---   &  ---     &    111.8 \\
 5506.778&Fe I&   0.9901  &  $-$2.797  &  102.8 &  ---     & ---   \\ 
 5339.928&Fe I&   3.266   &  $-$0.680  &   78.2 &  ---     & ---   \\ 
 5332.899&Fe I&   1.557   &  $-$2.940  &   72.2 &  ---     & ---    \\ 
 5324.178&Fe I&   3.211   &  $-$0.240  &  108.7 &  ---     &    117.1   \\
 5324.179&Fe I&   3.211   &  $-$0.103  &  ---   &  ---     &    117.1 \\
 5281.790&Fe I&   3.0380  &  $-$1.020  &   74.9 &  ---     &    93.0  \\
 5266.555&Fe I&   2.9980  &  $-$0.492  &  106.6 & 183.5    &   117.8  \\
 5254.955&Fe I&   0.1100  &  $-$4.764  &   79.9 &  ---     &   105.0   \\
 5250.210&Fe I&   0.1213  &  $-$4.940  &  ---   &  ---     &    85.0  \\
 5247.050&Fe I&   0.0872  &  $-$4.946  &  ---   & 198.1    &    88.4   \\
 5232.939&Fe I&   2.9400  &  $-$0.190  &  119.2 &  ---     &    134.4  \\
 5225.525&Fe I&   0.1101  &  $-$4.790  &  ---   & 224.3    &    91.6   \\
 5217.396&Fe I&   3.2100  &  $-$1.097  &   76.2 &  ---     & ---     \\
 5216.274&Fe I&   1.6080  &  $-$2.150  &  ---   &  ---     &   119.5  \\
 5202.340&Fe I&   2.1759  &  $-$1.838  &  ---   & 208.7    &   103.4     \\
 5198.710&Fe I&   2.2227  &  $-$2.134  &  ---   & 152.1    &    78.5   \\
 5194.940&Fe I&   1.5574  &  $-$2.090  &  ---   & 228.2    & ---    \\
 5192.343&Fe I&   2.9980  &  $-$0.521  &  108.4 &  ---     &   108.0   \\
 5171.595&Fe I&   1.4848  &  $-$1.793  &  133.7 &  ---     & 149.9    \\
 5166.281&Fe I&   0.0000  &  $-$4.190  &  119.6 &  ---     & ---    \\
 5006.119&Fe I&   2.8330  &  $-$0.610  &  ---   &  ---     &  115.3 \\
 4961.910&Fe I&   3.6344  &  $-$2.280  &  ---   &   54.9   & ---   \\
 4924.770&Fe I&   2.2786  &  $-$2.222  &  ---   &  156.8   & ---   \\
 4871.317&Fe I&   2.8650  &  $-$0.410  &  136.9 &  ---     & ---   \\
 4494.560&Fe I&   2.1973  &  $-$1.140  &  ---   &  ---     &   107.3 \\
 4484.219&Fe I&   3.6025  &  $-$0.720  &   60.0 &  ---     &    73.0   \\
 4466.551&Fe I&   2.8320  &  $-$0.590  &  122.9 &  ---     & ---    \\
 5316.615&Fe II&  3.1528  &  $-$1.850  &   94.7 &  ---     & ---    \\
 5276.002&Fe II&  3.1990  &  $-$1.940  &   76.6 &  ---     & ---     \\
 5234.625&Fe II&  3.2214  &  $-$2.050  &   61.5 &   78.3   &    80.5   \\
 4923.930&Fe II&  2.8910  &  $-$1.319  &  ---   &  160.3   &   130.2    \\
 4583.839&Fe II&  2.8070  &  $-$2.020  &   81.8 &  ---     &   110.3  \\
 4491.400&Fe II&  2.8555  &  $-$2.700  &  ---   &  ---     &     55.4 \\
\hline  

\end{tabular}
\end{table*}
}

{\footnotesize
\begin{table*}

{\bf Table B: Measured equivalent widths  }\\

\begin{tabular}{ c c c c c c  c }
  &   &   &    &   &  &   \\

\hline
W$_{lab}$&  ID & EP$_{low}$&log${gf}$& Eq widths (m\AA\,) & Eq widths (m\AA\,) &\\
          &     &          &         &HE~1305$+$0007& HD 5223  &   \\
\hline
  4571.096&Mg I &  0.0000  & $-$5.688  & 140.8 & 138.8$^{\rm as}$  &     \\
  5172.684&Mg I &  2.7117  & $-$0.402  & 250.7 & 281.0$^{\rm as}$  &     \\
  5528.405&Mg I &  4.340   & $-$0.620  & 171.3 & 186.6$^{\rm as}$ &     \\
  6102.727&Ca I &  1.880   & $-$0.793  &  74.0 & 91.5$^{\rm as}$   &     \\
  6122.226&Ca I &  1.890   & $-$0.316  & 125.9 &146.7$^{\rm as}$    &     \\
  5192.970&Ti I &  0.0211  & $-$0.950  & 138.9 & ---     &     \\
  4534.780&Ti I &  0.8360  &  0.336    & 119.4 & 131.0$^{\rm as}$   &     \\
  4805.090&Ti II&  2.0613  & $-$1.100  &  98.7 & 112.5$^{\rm as}$&     \\
  4798.507&Ti II&  1.0800  & $-$2.670  &  84.4 & 96.4$^{\rm as}$ &     \\
  5432.546&Mn I &  0.0000  & $-$3.795  &  31.4 & ---     &     \\
  5420.370&Mn I &  2.1427  & $-$1.460  &  43.0 & ---     &     \\
  6314.653&Ni I &  1.9355  & $-$1.770  &  45.8 & 59.3$^{\rm as}$ &    \\
  6111.060&Ni I &  4.0882  & $-$0.870  &  81.2 & 57.5$^{\rm as}$ &    \\
  4937.340&Ni I &  3.6060  & $-$0.390  & 124.5  & 162.6  &    \\
  4607.327& Sr I&  0.000   & $-$0.570  &  64.0  & 70.3   &     \\
  5200.406&Y II &  0.9923  & $-$0.570  &  74.2  & 90.2$^{\rm as}$&    \\
  5205.731&Y II &  1.0330  & $-$0.342  &  93.0  &103.9$^{\rm as}$&    \\
  6134.570&Zr I &  0.0000  & $-$1.280  &  30.1  & 19.7$^{\rm as}$&    \\
  6496.897&Ba II&  0.6040  & $-$0.369  & 379.0  &343.1$^{\rm as}$&     \\
  6141.727&Ba II&  0.704   & $-$0.069  & 390.0  &300.4  &     \\
  5853.668&Ba II&  0.6040  & $-$1.020  & 243.7 & 211.7$^{\rm as}$&     \\
  5188.206&La II&  2.4488  & $-$0.010  &  49.6 & 43.48$^{\rm as}$&     \\
  5259.380&La II&  0.1729  & $-$1.760  &  97.7 & 85.37$^{\rm as}$&      \\
  4773.941&Ce II&  0.9243  & $-$0.498  &  76.6 & 71.99  &      \\
  5512.062&Ce II&  1.0082  & $-$0.518  &  93.0 & 84.8$^{\rm as}$ &      \\
  5274.229&Ce II&  1.0444  & $-$0.323  & 103.0 & 100.7$^{\rm as}$&      \\
  5259.728&Pr II&  0.6334  &    0.080  & 110.6 &  96.4$^{\rm as}$&      \\
  5825.851&Nd II&  1.0807  & $-$0.760  &  76.9 & 48.3   &     \\
  5718.120&Nd II&  1.4100  & $-$0.510  &  80.0 & 63.1  &     \\
  5688.530&Nd II&  0.9860  & $-$0.250  &  90.2 & 75.1  &     \\
  5319.820&Nd II&  0.550   & $-$0.210  & 133.3 & 114.8 &    \\
  5293.169&Nd II&  0.8230  & $-$0.060  & 121.9 & 106.7$^{\rm as}$ &   \\
  5287.133&Nd II&  0.7446  & $-$1.300  &  60.2 &  38.5 &    \\
  5276.869&Nd II&  0.8590  & $-$0.610  &  86.4 &  64.3$^{\rm as}$ &    \\
  5255.506&Nd II&  0.2046  & $-$0.820  & 126.5 & 114.4 &    \\
  5249.590&Nd II&  0.9760  &  0.210    & 114.8 &  96.5 &    \\
  5212.361&Nd II&  0.2046  & $-$0.870  & 115.9 &  99.85$^{\rm as}$ &   \\
  4859.039&Nd II&  0.3200  & $-$0.830  & 104.7 & 102.0  &   \\
  4791.597&Sm II&  0.1000  & $-$1.846  &  65.2 &  42.4  &    \\
  4615.456&Sm II&  0.5400  & $-$1.262  &  83.0 &  67.8$^{\rm as}$  &   \\
\hline 
\end{tabular}
\\
``as" indicates lines are asymmetric \\
\end{table*}
}

Abundance anomalies particular to CH stars are generally explained as a
result of mass transfer from a companion star that has undergone its second
ascent of the giant branch, becoming an AGB star. Although such a scenario
seems to explain the observed behaviour in many classical CH stars,
including perhaps HE~1152$-$0355 and HD~5223, this cannot be generalized to
the case of HE~1305$+$0007, which exhibits a strong double enhancement of
both s-process elements (e.g., ${\rm [Pb/Fe]} = +2.37$) and r-process elements
(e.g., ${\rm [Eu/Fe]} = +1.97$). In the following, we focus on the abundance
pattern of this object for a detail discussion.

Figure 9 shows a comparison of the observed elemental abundance
distribution in HE~1305$+$0007, relative to Ba, with the solar abundance
patterns of the heavy elements associated with the r-process and s-process
solar abundances, taken from Burris et al. (2000). As can be seen from
inspection of this Figure, the observed $n$-capture abundances in
HE~1305$+$0007 do not agree with either of the scaled r-process or
s-process abundance patterns in the solar system; the general pattern is
clearly non-solar. The abundance pattern labelled as ``r+s'' is a simple
average of r-process and s-process solar abundances from Burris et
al. (2000); this pattern is closer to the observed abundance pattern in
HE~1305$+$0007 than either of the $r-$only and $s-$only patterns. While r+s
indicates a simple average, it should be possible to consider a weighted
average of the processes, where the weights are determined by relative
fractions of r- and s-process solar abundances that could perhaps better
reproduce the observed abundance pattern in HE~1305$+$0007.

The enhancements in Sr and Y are much lower than the enhancement in Ba,
consequently the ratios ${\rm [Y/Ba]}$ and ${\rm [Sr/Ba]}$ appear 
particularly low. The
elements with $57 \le $ Z $\le 62$ have abundance distributions closer to
the $r-$only curve, although La, Ce, and Pr are slightly above this curve,
while Nd and Sm are slightly below it. Eu and Pb lie between the values
from of the $r-$only and $s-$only solar-system curves, suggesting that the
observed pattern has contributions from {\it} both the $r-$ and $s-$
processes.

\begin{figure*}
\epsfxsize=10truecm
 \epsffile{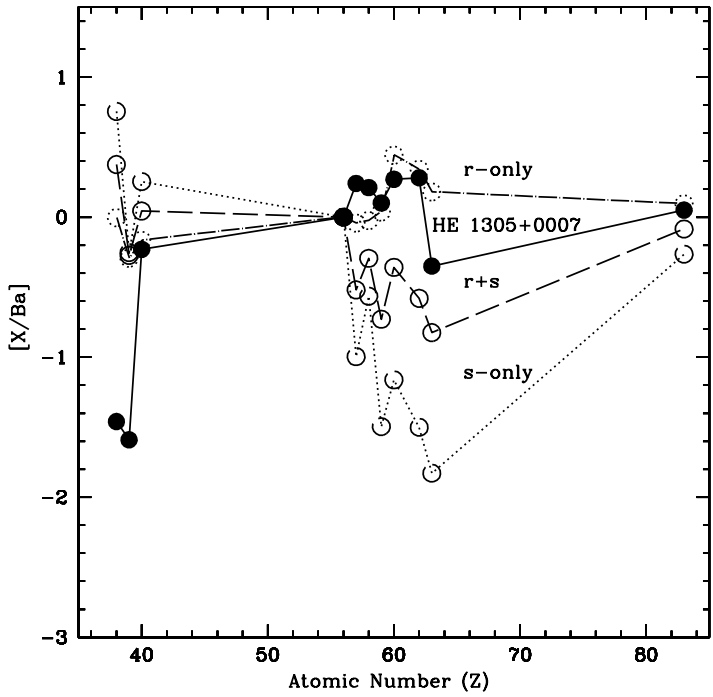}
\caption{The patterns of the heavy elements, relative to Ba, in
HE~1305$+$0007 are compared to the r-process and s-process solar abundances
from Burris et al. (2000).  The abundance pattern of HE~1305$+$0007 is
indicated by the solid line connecting the solid circles. The dotted line
indicates ``s-only'', the dot-dashed line indicates ``r-only'' and the dash
line indicates and average ``r+s'' abundance pattern connecting the open
circles. }
\label{Figure 9}
\end{figure*}

\begin{figure*}
\epsfxsize=10truecm
 \epsffile{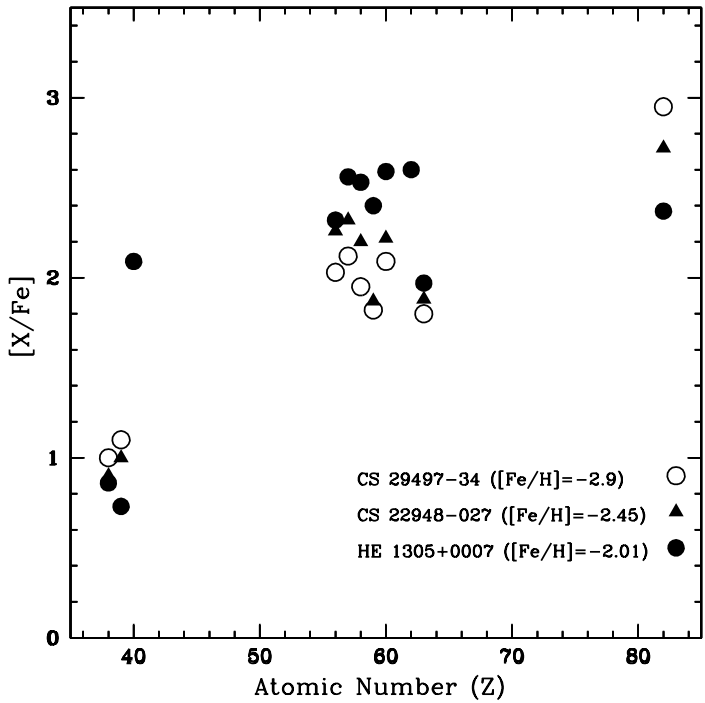}
\caption{A comparison of the distribution of $n$-capture elements
abundances ([X/Fe]) versus their atomic numbers (Z) in HE~1305$+$0007 with
those of two similar objects, CS~22948-027 and CS~29497-034. The solid
circles represent HE~1305$+$0007, the solid triangles represent
CS~22948-027, and the open circles represent CS~29497-034. }
\label{Figure 10}
\end{figure*}

\begin{figure*}
\epsfxsize=10truecm
 \epsffile{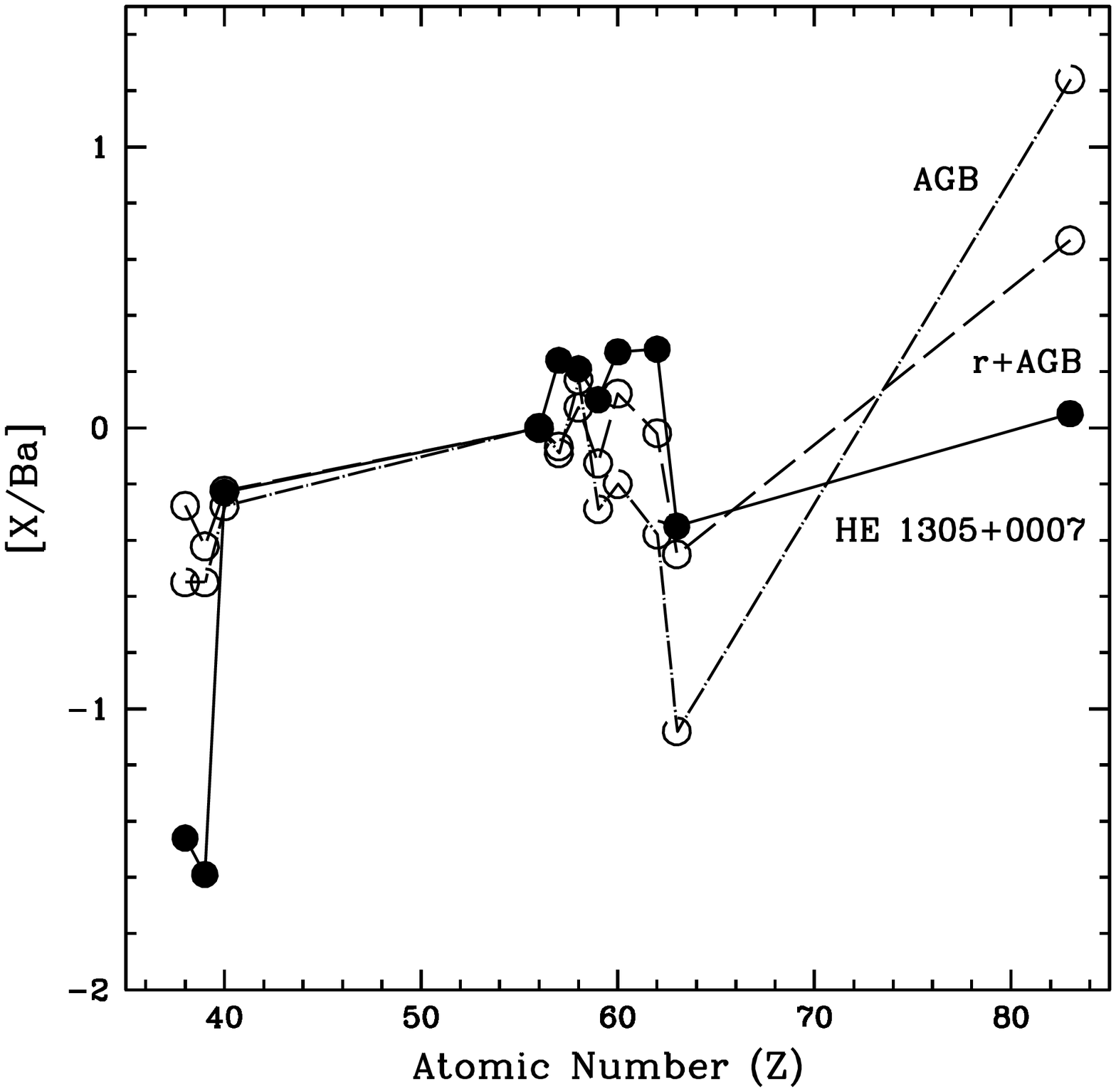}
\caption{The patterns of the heavy elements relative to Ba in
HE~1305$+$0007 (solid line, connecting the solid circles) are compared with
the abundances predicted to be found in the dredge-up material of a
low-metallicity AGB model (dot-dash line) from Goriely \& Mowlavi
(2000). The dashed line indicates the pattern from a simple mixture of the
abundances of the low-metallicity AGB model and the solar r-process
abundances. }
\label{Figure 11}
\end{figure*}

The number of very metal-poor stars that are known to exhibit double
enhancements of $r-$ and $s-$ process elements is still small. We consider
two examples, CS~22948-027 and CS~29497-034 (Hill et al. 2000, Barbuy et
al. 2005); these stars exhibit large enhancements of carbon as well as
atmospheric parameters, metallicities, and heavy-element abundance patterns
that are similar to HE~1305$+$0007. Hence, it is informative to compare the
abundance pattern of HE~1305$+$0007 with the abundance distribution of
these two stars. Radial velocity measurements show that both of the CS
stars are long-period binaries (Preston \& Sneden 2001, Barbuy et
al. 2005), indicating that mass accretion from the (now invisible)
more-massive companion occurred during its AGB phase (not RGB), explaining
the observed enhancements of the s-elements, including Pb (Barbuy et
al. 2005). Figure 10 shows the distribution of $n$-capture element
abundances ([X/Fe]) versus their atomic numbers (Z) for HE~1305$+$0007,
compared with those of CS~22948-027 and CS~29497-034. Table 7 summarizes
these abundances. We notice that these three stars exhibit overabundances
of the proton-capture element Na, which is believed to be formed by deep
CNO-burning in massive AGB stars. In general, enhancement of the
proton-capture elements Na and Al, although common in globular clusters, is
not observed in field metal-poor stars. Thus, the enhancement of Na in
these stars could also be an indication of pollution by an AGB
companion. From the similarity in abundance distributions, these three
stars seem to have originated from related nucleosynthetic and evolutionary
processes.  However, one should note that HE~1305$+$0007 has not yet been
shown to be a member of a binary system.

HE~1305$+$0007 also exhibits a large overabundance of the 3rd peak
s-process element Pb. In contrast to HE~1305$+$0007, all of the other known
metal-poor Pb-rich stars are not rich in r-process elements. In very
metal-poor AGBs, the s-process operates with a high ratio of neutrons to
seed nuclei that favours production of a high abundance of Pb. According to
Gallino et al. (1998), a large production of the doubly magic nucleus
$^{208}$Pb is due to AGB stars of low metallicity, which explains the
presence of Pb in low-metallicity stars. That the low-mass AGB stars (M
$\le$ 4M$_{\odot}$) play a dominant role in the production of $s-$elements
in the Galaxy was also supported by Travaglio et al. (2001).  Calculations
by Goriely \& Mowlawi (2000) and Goriely \& Siess (2001) also predict large
enhancements of third-peak $s-$elements relative to the 1st and 2nd
s-process peaks.

In Figure 11, we compare the abundances of HE~1305$+$0007 with the
abundances derived from a low-mass, low-metallicity AGB model (Goriely \&
Mowlavi 2000).  The abundance curve from the low-metallicity AGB model,
scaled to match the observed Ba abundance of HE~1305$+$0007, does not
appear to account for any of the stellar abundances of Sr, Y, Eu, let alone
Pb. In order to check for any additional r-process contributions, the
abundances of HE~1305$+$0007 are also compared with a simple mixture of the
abundances of the low-metallicity AGB model with solar r-process abundances
(scaled to match the [Ba/Eu] ratio in this star). The higher Eu abundance
compared with expectation from the pure s-process models suggests some
additional contribution from the r-process.  However, the observed
abundance pattern is clearly not adequately represented by a simple mixture
of $n$-capture processes.

{\footnotesize
\begin{table*}

{Table 7: Abundances of HE~1305$+$0007 compared with CS~22948-027 and CS~29497-034 }\\

\begin{tabular}{ l c c c c c}
  &   &   &    &    &   \\
\hline
Elements &  Z &  log\,${\epsilon}$ & [X/Fe]  & [X/Fe] & [X/Fe]  \\
  X     &     & Solar & HE~1305$+$0007 & CS~22948-027 & CS~29497-034\\ 
        &     &               & [Fe/H] = $-$2.0 & [Fe/H] = $-$2.45&[Fe/H] = $-$2.9 \\
        &     &               & [hs/ls] = 1.22 & [hs/ls] =  0.97&[hs/ls] =  1.28 \\
        &     &               & $^{12}$C/$^{13}$C = 10 & $^{12}$C/$^{13}$C = 14 &$^{12}$C/$^{13}$C= 12 \\
\hline
C I    & 6  & 8.39   & +1.84          &  +2.43    &  +2.63    \\
Na I   & 11 & 6.17   & +0.26 (D$_{2}$)&   +0.57   &  +1.18    \\
Na I   & 11 & 6.17   & +0.43 (D$_{1}$)&   +0.57   &  +1.18    \\
Ca I   & 20 & 6.31   & +0.13          &   +0.54   &  +0.45    \\
Sc II  & 21 & 3.05   & +0.09          &   ---     &  ---      \\
Ti I   & 22 & 4.90   & ---            &   +0.34   & +0.29     \\
Ti II  & 22 & 4.90   & +0.79          &   +0.54   & +0.44     \\
Mn I   & 25 & 5.39   & +0.14          &   ---     &  ---     \\
Ni I   & 28 & 6.23   & $-$0.25        & $-$0.01   & +0.01    \\
Sr I   & 38 & 2.92   & +0.86          &   ---     &  ---     \\
Sr II  & 38 & 2.92   & ---            & +0.90     & +1.00    \\
Y II   & 39 & 2.21   & +0.73          & +1.00     &+1.10     \\
Zr I   & 40 & 2.59   & +2.09          &   ---     &  ---     \\
Ba II  & 56 & 2.17   & +2.32          & +2.26     &+2.03     \\
La II  & 57 & 1.13   & +2.56          & +2.32     &+2.12     \\
Ce II  & 58 & 1.58   & +2.53          & +2.20     &+1.95     \\
Pr II  & 59 & 0.71   & +2.38          & +1.65     & +1.65    \\
Nd II  & 60 & 1.45   & +2.59          & +2.22     & +2.09   \\
Sm II  & 62 & 1.01   & +2.60          & +1.70     & +2.00   \\
Eu II  & 63 & 0.52   & +1.97          & +2.10     & +2.25   \\
Pb I   & 82 & 2.00   & +2.37          & +2.72     & +2.95    \\
\hline  

\end{tabular}
\end{table*}
}

\section{Conclusions}

A high-resolution spectroscopic analysis of the CEMP stars HE~1305$+$0007
and HE~1152$-$0355 indicates large enhancements of C and the s-process
elements relative to Fe. The 2nd-peak s-process elements are found to be
more enhanced than those of the 1st-peak s-process elements. HE~1305$+$0007
is also found to be characterised by a large enhancement of the 3rd-peak
s-process element lead (${\rm [Pb/Fe]} = +2.37$). The large enhancement of 
europium (${\rm [Eu/Fe]} = +1.97$) is another important feature of this star. 
The CEMP stars
exhibiting enhancements of Pb constitutes a growing group of 29 objects; we
raise this number to 30 by adding HE~1305$+$0007. In addition, this star is
likely to be one of a small number that exhibit double enhancement of
r-process and s-process elements, the CEMP-r/s group. The large
enhancements of $n$-capture elements exhibited by the classical CH star
HD~5223 is consistent with the abundance pattern generally noticed in other
CH stars. The enhancement of s-process elements in this star, along with
the large enhancement of carbon, indicate that a mass-transfer event in a
binary system from a companion AGB star that underwent s-process
nucleosynthesis during its lifetime. However, in HE~1305$+$0007, the
enhanced Eu abundance, together with the large abundance of the 3rd-peak
s-process element Pb, presents a challenge to understanding the formation
mechanism of this star. Among several scenarios proposed, the possibility
that a star with a high ratio of r-elements/Fe could, by accreting some
s-rich matter, become an $r/s$ star with a large [Eu/Fe], has been
discussed by many authors. A scenario invoking a triple system, one
component polluting the presently-observed star with r-elements, and
another polluting with s-elements has also been considered in the
literature, but was discarded on the grounds that such a triple system may
not be dynamically stable (Cohen et al. 2003). Qian \& Wasserburg (2003)
suggested that the double enhancement in the CEMP-r/s stars could be due to
the s-process occurring in an AGB member of a binary system, followed by
the r-process taking place in a subsequent accretion induced collapse (AIC)
of the white dwarf remnant of the former AGB.  However, the expected rate
of the occurence of AIC events in the Galaxy is low and highly
uncertain. This rarity cannot justify the substantial fraction of CEMP-r/s
stars ($\sim$ 30\%) observed among all the CEMP$-s$ stars (Bailyn \&
Grindlay 1990, Qian \& Wasserburg 2003).

Zijlstra (2004) proposed a different scenario, in which the primary
(evolved to an AGB star) transfers s-rich matter to the observed star, but
does not suffer large mass loss (owing to its low metallicity). At the end
of the AGB phase, the degenerate core explodes as an AGB
supernova. However, r-processing in such a scenario is not expected to be
significant.

A formation scenario for these stars involving high-mass AGB stars (8-10
M$_\odot$) was discussed at length by several authors (e.g., Barbuy et
al. 2005, Wanajo et al. 2006). The possibility of s-processing occurring in
a 10 solar mass star with an O-Ne-Mg core has been studied by Ritossa et
al. (1996); these authors demonstrated that the reaction
$^{22}$Ne(${\alpha}$, $n$)$^{25}$Mg is efficient in such high-mass stars,
owing to the high temperature 3$\times$10$^{8}$ K reached at the base of
the He-convective shell. An astrophysical scenario associated with the AGB
star in a binary system, in which the r-process might also occur, appears
to be a positive step toward understanding the formation mechanisms of the
CEMP-r/s stars. More quantitative studies of s-process nucleosynthesis in
this mass range is needed in order to make definitive conclusions. A
detailed and comprehensive discussion on several of the proposed formation
scenarios of CEMP-r/s stars is provided by Jonsell et al. (2006).

While in the present work we discuss the carbon and neutron-capture element
abundance distribution for only three CEMP stars, we intend to conduct
additional analyses for a larger sample of stars in the near future. Aoki
et al. (2006) provides one such study, including over 20 CEMP stars. Cohen
et al. (2006) describes a sample of 16 CEMP stars. An analysis of the
abundance patterns obtained from consideration of larger samples will
permit one to put the role of CEMP stars in the early history of Galactic
chemical evolution into better perspective. It is also important to seek
measurements of the crucial species Li, O, Na, and the mixing diagnostic
$^{12}$C/$^{13}$C ratio for these expanded samples of CEMP stars. In
addition to many theoretical propositions (Gallino et al. 1998, Goriely \&
Mowlavi 2000, Goriely \& Siess 2001), there now exists observational
evidence that suggests, in contrast to the $r-$only hypothesis of Truran
(1981), that the s-process could indeed operate even at very low
metallicities, as low as ${\rm [Fe/H]} = -3.1$ (Johnson \& Bolte 2002).  
Sivarani et al. (2006)
report on three unevolved (main-sequence turnoff) CEMP stars,
at least one of which (CS~29528-041) exhibits moderately high 
${\rm [Ba/Fe]}$ ($\sim$ +1.0) at a metallicity ${\rm [Fe/H]} = -3.3$. 
Masseron et al. (2006)
describe the remarkable discovery of an apparent thermally-pulsing AGB
star, CS~30322-023, which exhibits a clear s-process signature, at
 ${\rm [Fe/H]} = -3.5$. Obviously, there are many surprises yet to be 
revealed concerning
the operation of the s-process in the early Galaxy. Although uncertain, it
also seems likely that the impact of an early s-process on the chemical
evolution of carbon, nitrogen, oxygen, and the neutron-capture elements in
the Galaxy will critically depend on the Initial Mass Function of
early-generation stars (Abia et al. 2001; Lucatello et al. 2005; Karlsson
2006).\\

{\it Acknowledgements}\\
\noindent
This work made use of the SIMBAD astronomical database, operated at CDS,
Strasbourg, France, and the NASA ADS, USA. \\
T.C.B. acknowledges partial funding for this work from grant AST 04-06784,
as well as from grant PHY 02-16783: Physics Frontiers Center/Joint
Institute for Nuclear Astrophysics (JINA), both awarded by the
U.S. National Science Foundation. N.C. acknowledges financial support by
Deutche Forschungsgemeinschaft through grants Ch~214/3 and Re~353/44. He is
a research fellow of the Royal Swedish Academy of Sciences supported by a
grant from the Knut and Alice Wallenberg Foundation.  J.E.N. acknowledges
support from Australian Research Council grant DP0342613.\\
{}
\end{document}